\documentclass[aps,prb,reprint,superscriptaddress,showkeys,
floatfix,citeautoscript]{revtex4-1}

\usepackage[pdftex]{graphicx}
\usepackage{bm}
\usepackage{subfigure}
\usepackage{amsmath}
\usepackage{amsfonts}
\usepackage{amssymb}
\usepackage{epstopdf}
\usepackage{color}

\usepackage[normalem]{ulem}
\usepackage{hyperref}
\hypersetup{
   colorlinks,%
   citecolor=blue,%
   linkcolor=blue,%
   urlcolor=blue
}

\newcommand{\bA}{\mathbf{A}}
\newcommand{\bB}{\mathbf{B}}

\newcommand{\bdel}{\boldsymbol{\delta}}
\newcommand{\hs}{\mathrm{HS}}
\newcommand{\bk}{\mathbf{k}}
\newcommand{\bK}{\mathbf{K}}
\newcommand{\bq}{\mathbf{q}}
\newcommand{\br}{\mathbf{r}}
\newcommand{\bR}{\mathbf{R}}
\renewcommand{\Im}{\mathrm{Im}}
\newcommand{\pdag}{\phantom{\dag}}
\newcommand{\rme}{\mathrm{e}}
\newcommand{\sgn}{\mathrm{sgn}}
\newcommand{\tD}{\tilde{D}}
\newcommand{\tG}{\tilde{\Gamma}}
\newcommand{\ts}{\mathrm{TS}}

\newcommand{\tveps}{\tilde{\veps}}
\newcommand{\tW}{\tilde{W}}
\newcommand{\veps}{\varepsilon}

\begin{document}
\title{Sublattice symmetry breaking and Kondo-effect enhancement in
strained graphene}

\author{Dawei Zhai}
\affiliation{Department of Physics and Astronomy, and Ohio Materials Institute,
Ohio University, Athens, Ohio 45701-2979, USA}
\author{Kevin Ingersent}
\affiliation{Department of Physics, University of Florida,
P.O.\ Box 118440, Gainesville, Florida 32611-8440, USA}
\author{Sergio E. Ulloa}
\affiliation{Department of Physics and Astronomy, and Ohio Materials Institute,
Ohio University, Athens, Ohio 45701-2979, USA}
\author{Nancy Sandler}
\affiliation{Department of Physics and Astronomy, and Ohio Materials Institute,
Ohio University, Athens, Ohio 45701-2979, USA}

\date{\today}

\begin{abstract}
Kondo physics in doped monolayer graphene is predicted to exhibit unusual
features due to the linear vanishing of the pristine material's density of
states at the Dirac point. Despite several attempts, conclusive experimental
observation of the phenomenon remains elusive. One likely obstacle to
identification is a very small Kondo temperature scale $T_K$ in situations
where the chemical potential lies near the Dirac point. We propose tailored
mechanical deformations of monolayer graphene as a means of revealing unique
fingerprints of the Kondo effect. Inhomogeneous strains are known to produce
specific alternating changes in the local density of states (LDOS) away from
the Dirac point that signal sublattice symmetry breaking effects. Small LDOS
changes can be amplified in an exponential increase or decrease of $T_K$ for
magnetic impurities attached at different locations. We illustrate this behavior
in two deformation geometries: a circular ``bubble'' and a long fold, both
described by Gaussian displacement profiles. We calculate the LDOS changes for
modest strains and analyze the relevant Anderson impurity model describing a
magnetic atom adsorbed in either a ``top-site'' or a ``hollow-site''
configuration. Numerical renormalization-group solutions of the impurity model
suggest that higher expected $T_K$ values, combined with distinctive spatial
patterns under variation of the point of graphene attachment, make the top-site
configuration the more promising for experimental observation of signatures of
the Kondo effect. The strong strain sensitivity of $T_K$ may lift top-site Kondo
physics into the range experimentally accessible using local probes such as
scanning tunneling microscopy.
\end{abstract}

\maketitle

\section{Introduction}
\label{sec:intro}

The honeycomb structure of the graphene lattice has interesting consequences
for the low-energy electron dynamics. An effective massless dispersion near
the Dirac point, accompanied by spinor eigenstates with well-defined helicities
that impose specific phase relations between their components, yields high
carrier mobilities and unique optical properties for the pristine material
\cite{CastroNeto2009}. In addition, the strong $sp^2$ carbon bonding confers
remarkable mechanical properties that allow graphene to withstand high levels
of in-plane strain while being easily rippled under external stress, much
like paper \cite{Katsnelson}. The formation of wrinkles \cite{WChen2016,
SDeng2016,LYang2017}, folds \cite{KKim2011,HLim2015,YJiang2017}, and bubbles
\cite{JZabel2012} can be driven by lattice mismatch with a substrate
\cite{LGao2010,JXue2011}, intercalated impurities trapped during the deposition
process \cite{NLevy2010}, or directly by external application of controlled
stress fields \cite{CLau2012}. 

Such local deformations of graphene are responsible for inhomogeneous charge
density distributions with characteristics determined by the magnitude and
spatial dependence of the strain field. The connection between deformations
and charge inhomogeneities was quantitatively confirmed in recent measurements
of the local density of states (LDOS) via scanning tunneling microscopy (STM)
in setups with mobile (tip-induced) and static (intercalated impurity) local
deformations \cite{AGeorgi2017}. Analysis of STM images revealed local
sublattice-symmetry breaking in strained regions, whereby the two carbon atoms
within each unit cell are differentiated by contrasting signal intensities.
Interestingly, despite the local deformation, the gapless dispersion of the
pristine sample is maintained.
Strain-induced density enhancements have also been reported in transport
experiments through isolated folds, where charge confinement gives rise to
Coulomb-blockade features across the axis of the fold \cite{YWu2017}.

The studies cited in the previous paragraph suggest that strain may be used to
control local charge distributions and thereby reach regimes where
electron-electron interactions are important that are difficult to access in
undeformed graphene. An iconic example of strong correlations is the Kondo
effect, where mobile carriers collectively screen a localized magnetic moment
embedded in the system. This many-body phenomenon depends on the dynamics of
spin carriers and is sensitive to magnetic fields \cite{Hewson1997}. Its
characteristic energy scale, set by the Kondo temperature $T_K$, depends
strongly on both the hybridization matrix elements between localized and
delocalized levels and the LDOS of delocalized levels at the local-moment site.
Pristine graphene is predicted to be the setting for two distinct types of
Kondo physics. If the material is doped or gated so that its chemical potential
is away from the Dirac point, the Kondo effect is expected to be largely
conventional: the impurity contribution to bulk properties should show the same
dependences at sufficiently low temperatures, frequencies, and magnetic fields
as are found in a three-dimensional bulk metal \cite{Hewson1997}. In undoped
graphene, where the chemical potential lies precisely at the Dirac point,
theory instead predicts \cite{Sengupta2008,Zhu2010,Uchoa2011}
a ``pseudogap'' Kondo effect \cite{DWithoff1990} with
very different low-energy properties \cite{Bulla1997,CGonzalez-Buxton1998,%
PCornaglia2009,Uchoa2009,Uchoa2011,Li2013,LFritz2013,JJobst2013,Lo2014}.
In both the conventional and pseudogap cases, clear evidence for the Kondo
effect can be obtained only in experiments that are able to probe temperatures
below $T_K$. 

Kondo physics has recently been proposed to be the origin of features in
angle-resolved photoemission on Ce-intercalated graphene \cite{Hwang2018}.
However, the experimental setups most commonly pursued to realize the Kondo
effect in graphene involve either vacancies in the carbon lattice or adatoms
deposited on top of the sample. Claims of definitive detection of Kondo physics
in these settings remain controversial. Magnetotransport measurements on
irradiated (vacancy containing) graphene appear to reproduce the characteristic
temperature-dependence of the resistivity \cite{JHChen2011}, but doubt
has been cast on the Kondo intepretation by (i) the persistence of this
dependence as the chemical potential was tuned through the Dirac point
\cite{LFritz2013}, and (ii) the absence of Kondo signatures in the magnetic
response of iradiated graphene \cite{Nair2012} (though see \cite{JHChen2012}).
More recently, graphene with isolated vacancies has been reported to exhibit
Kondo features \cite{JMao2018} with a crucial dependence on curvature of the
graphene sheet \cite{DMay2018}.

Local STM probes of adatoms on graphene have yielded even more ambiguous
results. For example, early studies of cobalt adatoms on graphene found
features in the conductance expected for single and two-channel Kondo
effects, associating the two cases with different adsorption geometries
\cite{HManoharan2011}. However, similar features were later suggested to
arise instead from inelastic tunneling mediated by vibrations of cobalt
adatoms \cite{VBrar2011}. STM experiments involving hydrogen or fluorine
adsorbed on graphene have revealed no Kondo signatures \cite{Brihuega2016},
although different possible gating and/or doping regimes have not yet been
fully explored \footnote{I.\ Brihuega, private communication.}. 

First-principles prediction of the properties of adatoms on graphene has
also proved to be very challenging. An STM study of preferred adsorption
sites for nickel and cobalt adatoms on graphene with different substrate
conditions \cite{TEelbo2013} in some cases bore out, and in others
contradicted, the predictions of density-functional theory.
Theoretical analyses \cite{TWehling2010,Uchoa2009,HBXie2009} suggest that
the STM signatures of adatoms on graphene are highly sensitive to the
absorption geometry, which determines the relative energies of different
atomic orbitals, the effective Coulomb interactions between electrons in
various adatom orbitals, and the overlap integrals between adatom and host
orbitals. The Berry phase associated with the two inequivalent Dirac points
has also been predicted to play an essential role \cite{TWehling2010}.

It has been argued that part of the difficulty with observing the Kondo effect
with adatoms on graphene is the low density of states near the Dirac point,
which is expected to strongly suppress the Kondo temperature of the system
\cite{PCornaglia2009,Li2013,LFritz2013,JJobst2013} (especially for
cases of strict particle-hole symmetry, where no Kondo screening is possible
\cite{CGonzalez-Buxton1998,Li2013,LFritz2013}). Clear identification of Kondo
features may also be hindered by long-range charge fluctuations producing a
distribution of Kondo temperatures \cite{VMiranda2014} and by the
spatial delocalization of the impurity magnetic moment over nanometer scales
\cite{Brihuega2016}.

Recent experiments on graphene deposited on Ru(0001) surfaces \cite{JRen2014}
have highlighted strain as an important factor. Lattice mismatch with the
substrate imparts a rippled moir\'{e} superstructure to graphene. Cobalt
atoms were seen to adhere preferentially to graphene regions of high strain.
Fits of the differential conductance to Fano lineshapes suggested different
Kondo temperatures $T_K \simeq 12$\ K and $T_K \simeq 5$\ K
for adsorption at two types of site, each located at a local maximum of the
strain. The Kondo interpretation was supported by the observation of
magnetic-field-induced Zeeman splitting of the zero-bias conductance feature.
Although these results appear to provide strong evidence for Kondo physics,
it is hard to point to this as an example of Kondo screening by pristine
graphene since strong hybridization with Ru(0001) surface states washes out the
Dirac point and its linear dispersion \cite{Voloshina2016}.

We propose that with suitable modifications, experiments like those in Ref.\
\onlinecite{JRen2014} are very promising for the observation and
characterization of unique features of Kondo physics in graphene. The key idea
is to study samples in which the strained regions are not strongly hybridized
with a substrate. This may be accomplished by employing a substrate such as
hBN, or by focusing on free-standing graphene. In this paper, we show that
smooth deformations can induce modest modulations of the LDOS that lead to
strong changes in the Kondo temperature when the chemical potential lies in
the linear dispersion regime near, but not precisely at, the Dirac point.
The LDOS modulations consist of two components: one that breaks particle-hole
symmetry about the Dirac point, and one that breaks the symmetry between
sublattices $A$ and $B$. In certain regions near a deformation, an increase in
the LDOS of one sublattice is accompanied by a reduction of the LDOS at nearby
sites of the other sublattice. This local sublattice symmetry breaking is
amplified in the dependence of the Kondo temperature on the location at which
a magnetic atom adsorbs to the graphene host. In some cases, an exponential
enhancement of the Kondo scale will allow the observation of Kondo physics
where it would be undetectable in the absence of deformation.

We illustrate these ideas for two representative out-of-plane deformation
geometries: a localized Gaussian ``bubble'' with circular symmetry and an
extended Gaussian ``fold'' that preserves lattice translational symmetry along
the fold axis. We present and apply a formalism for calculating the graphene
LDOS changes resulting from modest strains, then analyze the relevant Anderson
impurity model describing a magnetic atom in one or other of the two most
probable adsorption configurations: so-called ``top'' and ``hollow'' sites.
Through nonperturbative numerical renormalization-group calculations, we
demonstrate that top-site adsorption above a single carbon atom leads to strong
strain sensitivity: even weak deformations (strain $\lesssim 1\%)$ can result
in enhancement of $T_K$ by at least an order of magnitude. For hollow-site
adsorption at the center of a carbon hexagon, it is unlikely that modest strains
can overcome a strong suppresssion of the Kondo scale in prsitine graphene that
results from destructive interference between tunneling of electrons between the
adatom and the six nearest host atoms.

The organization of the remainder of the paper is as follows.
Section \ref{sec:LDOS} reviews a description of strained graphene in terms
of scalar and (pseudo)vector gauge fields. This formalism is applied to compute
the LDOS near a Gaussian bubble and a Gaussian fold. Section
\ref{sec:Kondo} presents Anderson impurity models describing top-site
adsorption and hollow-site adsorption of a magnetic atom and emphasizes the
differing effects of strain in the two configurations. Numerical solutions of
the impurity model are used to map the variation of the Kondo temperature with
the location of top-site adsorption near a Gaussian bubble or a Gaussian fold.
Section \ref{sec:discuss} discusses  the results and presents suggestions for
experimental conditions favorable for the observation of the predicted features.

\section{LDOS of Strained Graphene}
\label{sec:LDOS}

A successful way to describe strain in graphene within a continuum Hamiltonian
formulation is by introducing effective (pseudo)gauge fields that change
electron dynamics without breaking time-reversal symmetry
\cite{Guinea2010,MVozmediano2010}.
Deformation-induced changes in the LDOS can be understood in terms of
(pseudo) Landau levels \cite{NLevy2010} or long-lived local resonances
(quasibound states) that are strain-field dependent \cite{YJiang2017,YWu2017}.
Due to the space inversion symmetry properties of the gauge fields,
strain is predicted to produce valley-filtered currents where electrons
near the two Dirac points are scattered differentially
\cite{MSettnes2016,AGeorgi2017}. These effects are expected to be enhanced
in the presence of external electromagnetic fields \cite{Milovanovic2017}.

This section reviews aspects of the continuum description of strained
graphene and presents calculations of the LDOS at points near out-of-plane
Gaussian deformations. Since such deformations have been the topic of
several previous studies\cite{Kusminskiy2011,Neek-Amal2012,
Ramon2014,MSchneider2015,Ramon2016,MSettnes2016}, Sec.\ \ref{subsec:fields}
presents a unified framework to enable comparison between various results.
The framework facilitates a discussion of various effects introduced by
strain and identifies those captured in scalar and pseudovector fields as
most relevant for electron dynamics in the energy range of interest. 
Section \ref{subsec:LDOS-changes} outlines the Green's function formalism
used to calculate the LDOS in deformed graphene. Based on underlying
lattice symmetries of graphene and the effective gauge fields, we derive
relations between deformation-induced changes in the LDOS in each valley and
on each sublattice. These relations point to the origins of the sublattice
symmetry breaking and particle-hole symmetry breaking that are evident in the
LDOS and, furthermore, are shown in Sec.\ \ref{sec:Kondo} to be magnified
in the spatial variation of the Kondo temperature scale.
The section concludes by illustrating the LDOS at different spatial
positions relative to a Gaussian bubble or fold, as calculated for several
representative combinations of model parameters

\subsection{Strain represented via effective gauge fields}
\label{subsec:fields}

We start with a model for undistorted monolayer graphene, with nearest-neighbor
bond length $a=1.42$\ \AA, assumed to lie in the plane $z=0$ with the $x$ [$y$]
axis chosen to point along one of the zigzag [bond] directions. Throughout
this paper, boldface symbols represent two-dimensional vectors in the
$x$-$y$ plane, and indices $i$, $j$ run over $1$ and $2$ (equivalent to $x$
and $y$, respectively).

Deformations of the two-dimensional graphene membrane that are smooth on
interatomic length scales can be described within continuum elasticity theory.
In-plane and out-of-plane displacements of carbon atoms from their equilibrium
positions are assumed to be described by functions $\mathbf{u}(\br)$ and
$h(\br)$, respectively, that vary slowly with undistorted in-plane position
$\br=(x,y)\equiv(r\cos\phi,\, r\sin\phi)$. To lowest order, the deformation is
described by an in-plane strain tensor \cite{BookElasticity}
\begin{equation}
\epsilon_{ij} =
  \frac{1}{2} (\partial_j u_i + \partial_i u_j
  + \partial_i h \partial_j h) .
\label{eq:strain}
\end{equation}

One effect of the deformation is to replace an undistorted nearest-neighbor
lattice vector $\bdel$ by a distorted counterpart
$\bdel'$ of length \cite{AKitt2012,AKitt2013}
\begin{equation}
\left|\bdel'\right|
= a + \frac{1}{a} \, \bdel\cdot\epsilon\cdot\bdel.
\end{equation}
As a result, the undistorted nearest-neighbor hopping matrix element
$t_0$ changes to \cite{VPereira2009}
\begin{align}
t
&= t_0 \exp\left[-\beta\left(|\bdel'|/a-1\right)\right] \notag \\
&\simeq t_0 [ 1 - (\beta / a^2) \bdel \cdot \epsilon
  \cdot \bdel ] ,
\label{eq:tapprox}
\end{align}
where $\beta\simeq 3$ is the Gr\"{u}neisen parameter \cite{ACroy2016}.

The modified hopping can be inserted into the the nearest-neighbor
tight-binding Hamiltonian describing the $\pi$ electrons in graphene.
Linearization around the Dirac points $K$ and $K'$ yields a low-energy
(effective) Hamiltonian for deformed graphene that, when written in the
valley-isotropic basis $(K_A,K_B)$ and $(-K'_B,K'_A)$
\cite{CBeenakker2008}, can be expressed in the form
\begin{equation}
H_{\tau} = v_F \boldsymbol{\sigma} \cdot
           [- i \hbar \boldsymbol{\nabla} - \tau\bK + \tau e\bA(\br)]
  + \Phi(\br) \, \sigma_0.
\label{eq:Hdef}
\end{equation}
Here, $\tau=1$ $(-1)$ identifies the $K$ ($K'$) valley centered around
wave vector $\tau\bK$ with $\bK=\left(4/3^{3/2},0\right)\pi/a$, $-e$ is
the electron charge, $v_F=3t_0 a / 2\hbar$ is the Fermi velocity, which
we take to be $v_F=10^6$ m/s (implying that $t_0 = 3.1$\ eV); $\sigma_x$
and $\sigma_y$ are Pauli matrices, and $\sigma_0$ is the $2\times 2$
identity matrix, all acting on the sublattice index. The effect of the
deformation is contained in effective gauge fields \cite{AndoCNT2002,%
MVozmediano2010}: a vector potential $\bA$ having components
\begin{equation}
A_x =\frac{\hbar\beta}{2ea} \, (\epsilon_{yy}-\epsilon_{xx}), \qquad
A_y = \frac{\hbar\beta}{ea} \, \epsilon_{xy},
\label{eq:A_x,y}
\end{equation}
and a scalar potential
\begin{equation}
\Phi(\br) = g_s ( \epsilon_{xx} + \epsilon_{yy}) .
\label{eq:Phi}
\end{equation}

Equation \eqref{eq:Hdef} takes the form of the Hamiltonian for free
electrons in the presence of an electric field
$\mathbf{E}=-\boldsymbol{\nabla} \Phi$ and a pseudomagnetic field
$\bB = \boldsymbol{\nabla} \times (\tau\bA)$. The pseudomagnetic field
changes signs between valleys, locally breaking the underlying inversion
symmetry of the honeycomb lattice but preserving time-reversal invariance.
This sign reversal gives $\tau\bA$ the character of a \textit{pseudovector}
gauge field.

The existence of a scalar potential of the form of Eq.\ \eqref{eq:Phi}
was originally argued \cite{AndoCNT2002} in the context of carbon
nanotubes, based on preservation of charge neutrality in a deformed area,
and led to an unambiguous prediction that $g_s>0$. The value of $g_s$
for graphene has been reported to be between 4\ eV and 6\ eV
\cite{MVozmediano2010,ACroy2016}. However, one well cited study corresponds
to $g_s = -2.5$\ eV \cite{Sloan2013} and another may be interpreted as
giving a similar value \cite{SMChoi2010}.
Given this uncertainty over the sign of $g_s$, below we illustrate
results obtained both for positive and negative values of $g_s$.

\begin{figure}[tb]
\centering
\includegraphics[width=3.38in]{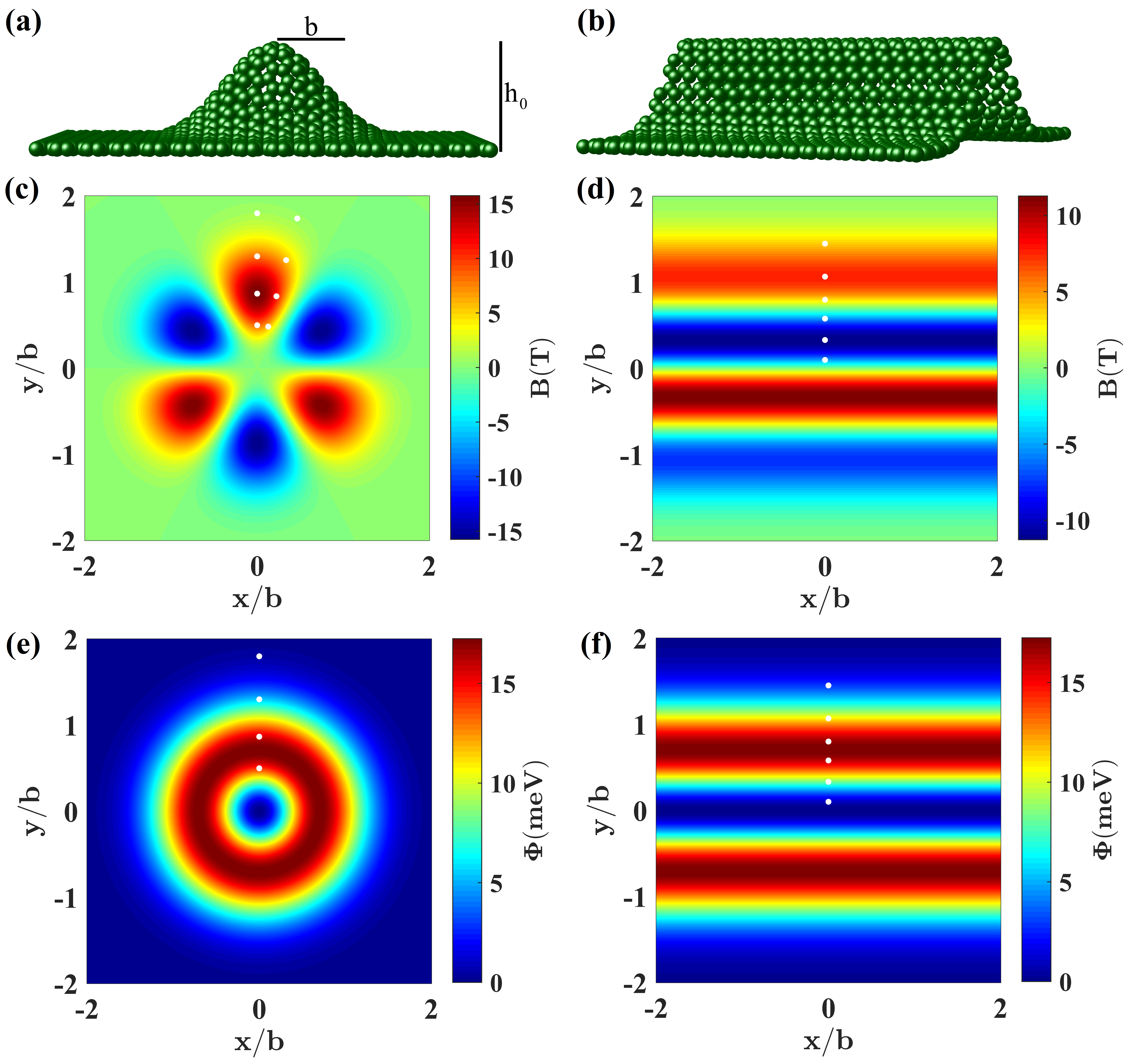}
\caption{\label{Fig1}
Two types of out-of-plane Gaussian deformation studied in this work:
(a) a circular bubble and (b) a long fold.
Below each schematic is a corresponding color map of (c), (d) the effective
magnetic field $B$ for electrons in the $K$ valley, and (e), (f) the scalar
potential $\Phi$ near both $K$ and $K'$. White circles in (c)--(f) indicate
positions where the LDOS is plotted in Figs.\ \ref{Fig2} and \ref{Fig3}.
Data shown are for deformations of peak height $h_0=1$\ nm and Gaussian
halfwidth $b=8$\ nm, with couplings $g_v=7$\ eV and $g_s=3$\ eV.
}
\end{figure} 

In this work, we consider setups where the deformations are of low aspect ratio,
i.e., the maximum out-of-plane displacement is much smaller than the in-plane
distance over which the deformation occurs. These conditions can be created,
for example, when atoms or molecules are intercalated between graphene and a
substrate, or through substrate engineering \cite{YJiang2017,Zhang2018}. In
such cases, in-plane atomic displacements $\mathbf{u}(\br)$ can be neglected
compared to out-of-plane displacements $h(\br)$ in Eq.\ \eqref{eq:strain}
\cite{Katsnelson}.

We focus on two specific deformation geometries:
a circular ``bubble'' and a long ``fold'' extending along the $x$ (zigzag)
direction, both having Gaussian out-of-plane height profiles
\cite{Neek-Amal2012,Ramon2014,MSchneider2015,Ramon2016,MSettnes2016}.
The bubble, shown schematically in Fig.\ \ref{Fig1}(a), is a centrosymmetric
deformation described in the Monge parametrization \cite{DNelsonbook} by a
height function
\begin{equation}
h(\br) = h_0 \, \rme^{-r^2/b^2} ,
\label{eq:h:bubble}
\end{equation}
while the long fold, sketched in Fig.\ \ref{Fig1}(b), has an out-of-plane
profile
\begin{equation}
h(\br) = h_0 \, \rme^{-y^2/b^2} .
\label{eq:h:fold}
\end{equation}
For these specific deformations, the description in terms of continuum
elasticity theory is valid as long as we take $\eta = (h_0 / b)^2 \ll 1$
\cite{BookElasticity}, while the condition $b \gg a$ ensures the absence
of inter-valley scattering. Under both geometries, the strain field given
by Eq.\ \eqref{eq:strain} is spatially inhomogeneous and has a peak magnitude
$\epsilon_{\max}=\eta^2/\rme$ with $\rme$ being Euler's number (not to be
confused with the elementary positive charge $e$). All results presented in
this paper are for deformations with a peak height $h_0=1$\ nm and a Gaussian
halfwidth $b=8$\ nm, for which the maximum strain takes a rather small value
$\epsilon_{\max}\simeq 0.6\%$.

For the circular bubble, Eq.\ \eqref{eq:A_x,y} predicts a vector potential
\begin{equation}
\bA(\br) = \frac{g_v \eta^2}{e v_F} f\left(\frac{r}{b}\right) \,
  ( -\cos2\phi, \; \sin2\phi ) ,
\end{equation}
where $g_v=\hbar\beta v_F/2a \simeq 7$\ eV and $f(z)=2z^2\exp(-2z^2)$,
while Eq.\ \eqref{eq:Phi} gives a scalar potential
\begin{equation}
\Phi(\br) = g_s \eta^2 f\left(\frac{r}{b} \right) .
\label{eq:Phi:bubble}
\end{equation}
The corresponding gauge fields for the long fold are
\begin{equation}
\bA(\br) = \frac{g_v \eta^2}{e v_F} f\left(\frac{y}{b}\right) \,
	(1, \; 0) ,
\end{equation}
and
\begin{equation}
\Phi(\br) = g_s \eta^2 f\left(\frac{y}{b}\right) .
\label{eq:Phi:fold}
\end{equation}

Figures \ref{Fig1}(c) and \ref{Fig1}(d) map the effective out-of-plane
magnetic field $B(\br)=\nabla\times\bA(\br)=B(\br)\,\hat{\mathbf{z}}$ as
experienced by electrons in the $K$ valley. This field has the opposite
sign for electrons in the $K'$ valley. The effective field $B(\br)$
produced by a circular bubble [Fig.\ \ref{Fig1}(c)] exhibits three-fold
rotational symmetry about the deformation peak at $\br=\mathbf{0}$ with
alternating positive and negative ``petals,'' as previously reported
\cite{MVozmediano2010,Neek-Amal2012,Ramon2014,AGeorgi2017}. The
effective magnetic field created by a long fold [Fig.\ \ref{Fig1}(d)]
is odd under $y\to-y$ with alternating positive and negative regions
on each side of the fold axis. By contrast, the corresponding scalar
fields, mapped in Figs.\ \ref{Fig1}(e) and \ref{Fig1}(f), exhibit the
even symmetry of the deformation profile.

Besides the gauge fields described above, which originate in
bond-length deformations, strain induces other effects:
(1) Additional gauge fields arise from changes in the orientation of $\pi$
orbitals as the graphene membrane is displaced out of the plane
\cite{CKane1997,EAKim2008}. These fields can be shown to be smaller than
those in Eqs.\ \eqref{eq:A_x,y} and \eqref{eq:Phi} by a multiplicative
factor of order $(a/b)^2$.
(2) Shifts in the positions of the $K$ and $K'$ points in reciprocal space
lead to renormalization of the effective Fermi velocity $v_{F}$
\cite{MVozmediano2010,SMChoi2010,deJuan2012,GNaumis2014,OlivaLeyva2015,%
OlivaLeyva2018}. Based on Ref.\ \onlinecite{SMChoi2010}, we estimate that
this renormalization induces fractional LDOS changes
$|\Delta\rho|/\rho \lesssim 4 \epsilon_{\max}$. For the specific situations
illustrated in this paper, where $(a/b)^2 \simeq 0.02$ and
$\epsilon_{\max} = 0.6\%$, effects (1) and (2) can be safely neglected.

\subsection{LDOS changes induced by strains}
\label{subsec:LDOS-changes}

In the continuum limit, the LDOS at position $\br$ and energy $E$ of
electrons on sublattice $\ell=1$ or $2$ in valley $\tau=\pm 1$ is
\begin{equation}
\rho_{\tau,\ell}(\br,E)
  = -\pi^{-1} s_E \, \Im \left[G_{\tau}(\br,\br,E)\right]_{\ell\ell}.
\label{eq:rho_tau,ell:def}
\end{equation}
Here, $s_E=\sgn\, E$ and $[G_{\tau}(\br,\br',E)]_{\ell\ell'}$, an
element of the $2\times 2$-matrix single-particle Green's function
\begin{align}
G_{\tau}(\br,\br',E)
&= \langle\br|(E+i s_E \,0^+ - H_{\tau})^{-1}|\br'\rangle ,
\label{eq:G_tau:def} \\
&\equiv e^{i\tau\bK\cdot(\br-\br')} \bar{G}_{\tau}(\br,\br',E) \notag
\end{align}
describes propagation of an electron in valley $\tau$ from spatial
location $\br'$ in sublattice $\ell'$  to location $\br$ in sublattice
$\ell$. The distinction between $G_{\tau}$ and $\bar{G}_{\tau}$,
which is usually neglected and does not affect the density of states [Eq.\
\eqref{eq:rho_tau,ell:def}], will prove to be important when we consider
hollow-site adsorption of a magnetic impurity (see Sec.\
\ref{subsec:hollow}).

In the limit $b\gg a$ considered in this work, the slowly varying
deformation induces negligible inter-valley scattering and one can calculate
$G_{\tau}(\br,\br',E)$ perturbatively in the Born approximation
\cite{BookGreensFunction} as
\begin{align}
&G_{\tau}(\br,\br',E) \simeq G_{0,\tau}(\br,\br',E) \notag \\
& \;\; +\int G_{0,\tau}(\br,\br_1,E) \, [V_{\tau}(\br_1) +
\Phi(\br_1) \sigma_0] \, G_{0,\tau}(\br_1,\br',E) \, d\br_1 ,
\label{eq:G_tau:Born}
\end{align}
where $G_{0,\tau}$ is the Green's function of pristine graphene, and
the first and second terms in the square brackets describe scattering
at a location $\br_1=(x_1,\, y_1)\equiv (r_1\cos\phi_1,\, r_1\sin\phi_1)$
due to the effective vector and scalar potentials, respectively. When written
in the valley-isotropic basis $(K_A,\, K_B)$ and $(-K'_B,\, K'_A)$,
$\bar{G}_{0,\tau}\equiv\bar{G}_0$ is independent of $\tau$. Within the
approximation of a linear dispersion in pristine graphene, i.e.,
$\veps_{\bk}=\pm\hbar v_F |\bk-\tau\bK|$ (valid for $|\veps_{\bk}|\ll t_0$),
one can show \cite{Ferreira2011} that
\begin{equation}
\bar{G}_0(\br,\br',E) = -\frac{iq}{4\hbar v_F} 
\begin{pmatrix}
s_E \, H_0(qd) & i \rme^{-i\phi_d} H_1(qd) \\[1ex]
i \rme^{i\phi_d} H_1(qd) & s_E \, H_0(qd) .
\end{pmatrix}
\label{eq:Gbar_0}
\end{equation}
Here, $q=|E|/\hbar v_F$, $H_n(x)$ is the order-$n$ Hankel function of the
first kind, and $\mathbf{d}=\br-\br'\equiv (d\cos\phi_d, d\sin\phi_d)$.
After summation over the valley index, the pristine Green's function leads
to a pristine density of states per sublattice, per spin orientation, and
per unit area
\begin{equation}
\rho_0(E) = \frac{|E|}{2\pi\hbar^2 v_F^2} .
\label{eq:rho_0:lin}
\end{equation}

For the Gaussian bubble, the scattering matrix arising from the pseudovector
field $\tau\bA(\br)$ is
\begin{align}
V_{\tau}(\br_1)
&= \tau e v_F \boldsymbol{\sigma} \cdot \bA (\br_1) \notag \\
&= -\tau g_v \eta^2 f\left(\frac{r_1}{b}\right)
\begin{pmatrix}
0 & \rme^{i2\phi_1}\\
\rme^{-i2\phi_1} & 0
\end{pmatrix} .
\label{eq:V:bubble}
\end{align}
The corresponding quantity for the Gaussian fold is
\begin{equation}
V_{\tau}(\br_1)=\tau g_v \eta^2 f\left(\frac{y_1}{b}\right)
\begin{pmatrix}
0 & 1\\
1 & 0
\end{pmatrix}.
\label{eq:V:fold}
\end{equation}

Using Eqs.\ \eqref{eq:Phi:bubble}, \eqref{eq:Phi:fold}, \eqref{eq:G_tau:Born},
\eqref{eq:Gbar_0}, \eqref{eq:V:bubble}, and \eqref{eq:V:fold}, and noting that
the elements of the unperturbed Green's function satisfy
\begin{align}
[G_0(\br,\br',E)]_{j\jmath'}
&= (-1)^{j-\jmath'+1} [G_0(\br,\br',-E)]_{j\jmath'} \notag \\[-1.75ex]
\\[-1.75ex]
&= (-1)^{j-\jmath'} [G_0(\br',\br,E)]_{j\jmath'} \notag
\end{align}
for $j,\,\jmath' \in \{1,\,2\}$, one can show that
$\Delta G_{\tau}^v(\br,\br_1,E)$ and
$\Delta G_{\tau}^s(\br,\br_1,E)$---respectively the pseudovector and scalar
contributions to the integral in Eq.\ \eqref{eq:G_tau:Born}---when
evaluated at $\br'=\br$, satisfy
\begin{alignat}{3}
[\Delta G_{\tau}^s]_{11}
&=[\Delta G_{\tau}^s]_{22}
&&=[\Delta G_{-\tau}^s]_{11}
&&=\text{even in $E$}, \\
[\Delta G_{\tau}^v]_{11}
&=-[\Delta G_{\tau}^v]_{22}
&&=-[\Delta G_{-\tau}^v]_{11}
&&=\text{odd in $E$}.
\end{alignat}
Taking into account the ordering of the basis in each valley,
Eq.\ \eqref{eq:rho_tau,ell:def} yields corresponding
deformation-induced shifts in the local density of states that satisfy
\begin{alignat}{2}
\label{eq:Drho_s:symm}
\Delta\rho_{K,A}^s(\br,E)
&=-\Delta\rho_{K,A}^s(\br,-E)
&&=\Delta\rho_{K',A}^s(\br,E) \notag \\[-1.75ex]
\\[-1.75ex]
&=\Delta\rho_{K,B}^s(\br,E)
&&=\Delta\rho_{K',B}^s(\br,E) , \notag \\
\label{eq:Drho_v:symm}
\Delta\rho_{K,A}^v(\br,E)
&=\Delta\rho_{K,A}^v(\br,-E)
&&=\Delta\rho_{K',A}^v(\br,E) \notag \\[-1.75ex]
\\[-1.75ex]
&=-\Delta\rho_{K,B}^v(\br,E)
&&=-\Delta\rho_{K',B}^v(\br,E) . \notag
\end{alignat}
In summary, valleys $K$ and $K'$ contribute equally to the net change of
LDOS $\Delta\rho_{\ell}(\br,E)$ experienced by each sublattice. While the
contribution of the scalar potential to $\Delta\rho_{\ell}(\br,E)$ is
identical for the two sublattices but odd in energy $E$, the shift coming
from the vector potential is even in $E$ but has opposite signs for
$\ell=A$ and $\ell=B$. The last property will prove to be the origin of
sublattice symmetry breaking in signatures of Kondo physics.

In the following sections, we present sublattice-resolved LDOS shifts
$\Delta\rho_{\ell}^{\alpha}(\br,E)$ ($\alpha = v,\, s$) and the total LDOS
$\rho_{\ell}(\br,E)$. The LDOS shifts are calculated via the method described
above, numerically integrating Eq.\ \eqref{eq:G_tau:Born} using the linearized
approximation [Eq.\ \eqref{eq:Gbar_0}] for $\bar{G}_0(\br,\br',E)$. The full
LDOS is computed as
\begin{equation}
\rho_{\ell}(\br,E)
  = \rho_0(E)+\sum_{\alpha=s,v}\Delta\rho_{\ell}^{\alpha}(\br,E),
\label{eq:LDOS:full}
\end{equation}
where $\rho_0(E)$ is the \emph{exact} nearest-neighbor tight-binding density of
states of pristine graphene \cite{CastroNeto2009}. The use of the exact
$\rho_0(E)$ makes little difference on the energy scales $|E|\ll t_0$ spanned by
Figs.\ \ref{Fig2} and \ref{Fig3} but it allows for a more realistic treatment
of higher energy scales, important for an accurate computation of the Kondo
temperature.

\subsection{LDOS for graphene with a Gaussian bubble deformation}
\label{subsec:LDOS-bubble}

\begin{figure*}[t]
\centering
\includegraphics[width=5.0in]{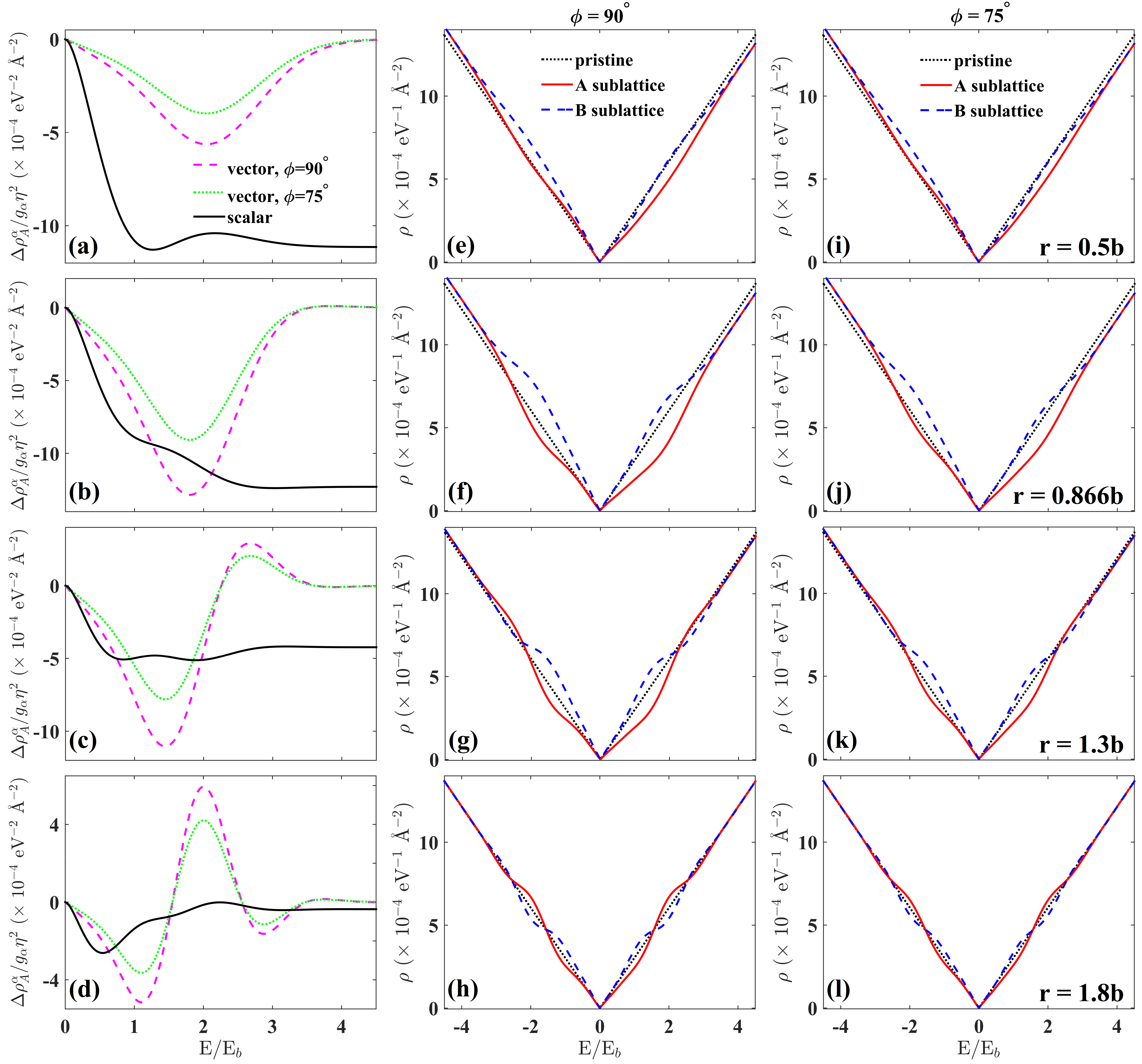}
\caption{\label{Fig2}%
LDOS near a Gaussian bubble. Each row shows data at a different radial
distance $r$ from the center of the deformation (labeled in the right
panel) and at two polar angles, $\phi=90^{\circ}$ and $75^{\circ}$; see
the locations marked by white dots in Fig.\ \ref{Fig1}(c).
(a)--(d) Scaled changes $\Delta\rho_A^{\alpha}/(g_{\alpha}\eta^2)$ in the
$A$-sublattice LDOS (valley-summed, per spin orientation, per unit area)
due to the scalar ($\alpha=s$, solid lines) and vector ($\alpha=v$,
dashed and dotted lines) gauge fields, plotted vs positive energy $E$
divided by $E_b=\hbar v_F/b$. (Changes due to the scalar potential are
independent of $\phi$.)  Note the different vertical scales in these panels.
(e)--(h) LDOS (valley-summed, per spin orientation, per unit area) for the
$A$ (solid lines) and $B$ (dashed lines) sublattices at locations having
coordinates $\phi=90^{\circ}$ and the radii $r$ used in (a)--(d),
respectively. The LDOS of pristine graphene is shown for reference (dotted
lines).
(i)--(l) Same as (e)--(h) except for locations at $\phi=75^{\circ}$.
Data in (e)--(l) were calculated for $h_0=1$\ nm, $b=8$\ nm, $g_v = 7$\ eV,
and $g_s = 3$\ eV.
}
\end{figure*} 

Figure \ref{Fig2} contains representative results for the graphene LDOS in the
vicinity of a Gaussian bubble deformation. Data are shown at four different
distances ($r=0.5b$, $0.866b$, $1.3b$, and $1.8b$) from the center of the
bubble along two different directions: $\phi=90^{\circ}$ (along a carbon-carbon
bond direction, which forms the symmetry axis of one of the petals in the
pseudomagnetic field) and $\phi=75^{\circ}$ (halfway in angle between the
petal symmetry axis  and a direction $\phi=60^{\circ}$ of zero pseudomagnetic
field). The eight chosen locations are marked by white circles in Fig.\
\ref{Fig1}(c). Along the direction $\phi=90^{\circ}$, $r=0.866b$ is a position
of maximum pseudomagnetic field $B$. At each $r$ value, rotating from
$\phi=90^{\circ}$ to $\phi=75^{\circ}$ moves off the petal symmetry axis,
resulting in a reduction in $B$.

Panels (a)--(d) in Fig.\ \ref{Fig2} show contributions to the change in the
LDOS (valley-summed, per spin orientation, per unit area) for sublattice $A$
due to the vector ($\alpha=v$) and scalar ($\alpha=s$) gauge fields, plotted as
$\Delta\rho_A^{\alpha}/(g_{\alpha}\eta^2)$ vs $E/E_b$, where $E_b=\hbar v_F/b$.
Here, $g_{\alpha}\eta^2$ determines the maximum magnitude of the $\alpha$
field, while $E_b$ is the natural energy scale associated with spatial
variations over a length $b$. When scaled in this fashion, the LDOS changes at
given $r/b$ and $\phi$ are universal functions, independent of the deformation
length scales ($h_0$ and $b$) as well as the gauge couplings ($g_{\alpha}$).
The results can be extended to negative values of $E$ and to the $B$ subattice
using the symmetry relations in Eqs.\ \eqref{eq:Drho_s:symm} and
\eqref{eq:Drho_v:symm}.

At each of the eight locations shown in Figs.\ \ref{Fig2}(a)--\ref{Fig2}(d), the
vector LDOS shift $|\Delta\rho_A^v(\br,E)|$ increases from zero at $E=0$,
passes through one or more maxima at energies $E\sim O(E_b)$, and then decreases
toward zero for $E\gg E_b$. The oscillations on the energy scale $E_b$ are the
result of interference between scattering at different locations throughout
the deformed region.
The greatest value of $|\Delta\rho_A^v(\br,E)|$ over all $E$ correlates closely
with the magnitude of the pseudomagnetic field $B(\br)$. The scalar shift
$|\Delta\rho_A^s(\br,E)|$ rises from zero at $E=0$, exhibits interference
features around $E=O(E_b)$, and saturates for $E\gg E_b$ at a value proportional
to $\Phi(\br)$. This saturation behavior has a simple interpretation: for
$E\gg E_b$, electrons experience an energy shift equal to the local scalar
potential $\Phi(\br)$, resulting in a LDOS shift $\Delta\rho_A^s(\br,E) = 
\rho_0(E-\Phi(\br))-\rho_0(E)\simeq -\Phi(\br) / (2\pi\hbar^2 v_F^2)$.

The functional form of the LDOS changes can be determined analytically for
$|E|\ll E_b$. Due to the exponential decay of the scattering potentials
$V_{\tau}(\br_1)$ and $\Phi(\br_1)$ for $|\br_1|\gg b$, the integral over
$\br_1$ in Eq.\ \eqref{eq:G_tau:Born} can be restricted to values of $|\br_1|$
smaller than a few times $b$. Then the argument of the Hankel functions in
Eq.\ \eqref{eq:Gbar_0}, $k d\equiv (E/E_b)|\br_1 - \br|/b$, vanishes
as $E/E_b\to 0$. Using the forms of the Hankel functions for small
arguments, one can deduce that  $\Delta\rho_A^v(\br,E)\propto |E/E_b|$ and
$\Delta\rho_A^s(\br,E)\propto s_E (E/E_b)^2 \ln|E/E_b|$, relations that are
in good agreement with our numerical data for $|E|\lesssim 0.1 E_b$ for
positions inside the deformed region. The corresponding analysis for positions
outside this region results in a leading order contribution from the vector
potential proportional to $(E/E_b)^4 e^{-|E/E_b|^2}$ plus a term due to the
scalar potential that behaves as $s_E (E/E_b)^2 \ln|E/E_b|$.

The remaining panels in Fig.\ \ref{Fig2} plot the full LDOS $\rho_A(\br,E)$
(solid line) and $\rho_B(\br,E)$ (dashed line) vs $E/E_b$ for our reference
case of a Gaussian bubble with a maximum height $h_0=1$\ nm and a halfwidth
$b=8$\ nm ($E_b \simeq 0.082$\ eV), with couplings $g_v=7$\ eV and
$g_s=3$\ eV. Panels (e)--(h) show results for $\phi=90^{\circ}$ at the same
$r$ values as in (a)--(d), respectively, while (i)--(l) represent
$\phi=75^{\circ}$. Each panel includes for comparison the linear LDOS of
pristine graphene (dotted line). These plots clearly show the shift in
spectral weight from $E>0$ to $E<0$ induced by the scalar potential $\Phi$,
as well as the spectral weight transfer between the two sublattices that
arises from the pseudovector potential $\tau\bA$. At each location $\br$, the
greatest difference between the $A$ and $B$ sublattice LDOS occurs for energies
$E\sim E_b$, while the greatest difference between the energy-integrated LDOS
on the two sublattices occurs at the position $r=0.866b$, $\phi=90^{\circ}$ of
strongest pseudomagnetic field.

\subsection {LDOS near a long Gaussian fold deformation}
\label{subsec:LDOS-fold}

\begin{figure}
\includegraphics[width=3.38in]{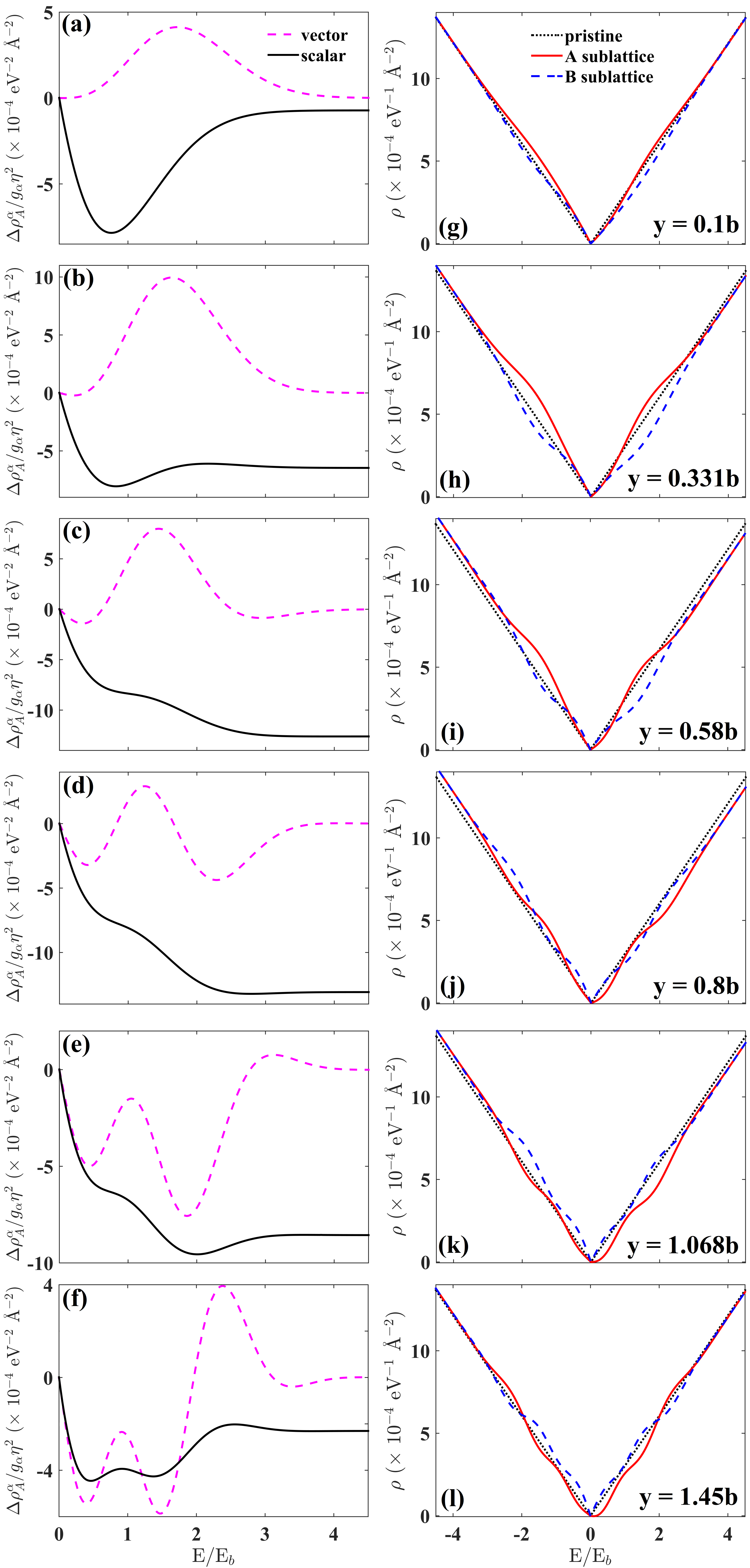}
\caption{\label{Fig3}
LDOS near a long Gaussian fold. Each row shows data at a different
perpendicular distance $y$ from the central axis of the deformation
(labeled in the right panel); see the locations marked by white dots in
Fig.\ \ref{Fig1}(c).
(a)--(f) Scaled changes $\Delta\rho_A^{\alpha}/(g_{\alpha}\eta^2)$ in the
$A$-sublattice LDOS (valley-summed, per spin orientation, per unit area)
due to the scalar ($\alpha=s$, solid lines) and vector ($\alpha=v$,
dashed lines) gauge fields, plotted vs positive energy $E$ divided by
$E_b=\hbar v_F/b$.
(g)--(l) LDOS (valley-summed, per spin orientation, per unit area) for the
$A$ (solid lines) and $B$ (dashed lines) sublattices at the locations
considered in (a)--(f), respectively, calculated for $h_0=1$\ nm, $b=8$\ nm, 
$g_v = 7$\ eV, and $g_s = 3$\ eV. The LDOS of pristine graphene is shown
for reference (dotted lines).
}
\end{figure} 

Figure \ref{Fig3} shows the LDOS (valley-summed, per spin orientation, per unit
area) near an extended Gaussian fold deformation. Results are presented for the
six locations marked by white circles in Figs.\ \ref{Fig1}(d) and \ref{Fig1}(f),
chosen to sample the range of coordinates $y$ (measured from the fold axis
$y=0$) over which the pseudomagnetic field and the scalar deformation potential
exhibit their strongest spatial variations. Panels (a)--(f) plot changes in the
$A$-sublattice LDOS due to the vector and scalar gauge fields, scaled in the
same way as the results in the left column of Fig.\ \ref{Fig2}. Just as for the
bubble, $\Delta\rho_A^v/(g_v\eta^2)$ is largest at the locations of greatest
magnitude of the pseudomagnetic field, which here are $y=0.331b$, [panel (b)]
and $y=1.068b$ [panel (e)]. Similarly, $\Delta\rho_A^s/(g_s\eta^2)$ is largest
at the peak location $y/b=2^{-1/2}\simeq 0.7$ of the scalar potential [not far
from the positions shown in panels (c) and (d)]. Larger $y$ values are associated
with increasing numbers of features in the $\Delta\rho_A^v(\br,E)$ around energy
scales of order $E_b$, resulting from extended regions of interference as
electrons scattering from the regions of largest pseudomagnetic field and scalar
potential must travel longer path lengths en route to locations $\br$ further
from the deformation axis. 

Analysis of the low-energy asymptotic behavior of LDOS shifts is more complicated
than in the case of the Gaussian bubble because (as noted above), the scattering
region is unbounded along the $x$ direction, allowing the arguments of the
Hankel functions entering Eq. \eqref{eq:Gbar_0} to take arbitrarily large values
for any $|E| \ll E_b$. For positions $|y| \gtrsim b$ both the vector and scalar
LDOS shifts are well described for $|E|\lesssim 0.1 E_b$ by a leading term
proportional to $|E/E_b|$. The LDOS also exhibit oscillations at energies
$E \gtrsim E_b$, similar to those shown by the bubble, that can be traced back
to interference between scattering at different locations throughout the deformed
region.

Panels (g)--(l) in Fig.\ \ref{Fig3} plot the full LDOS $\rho_A(\br,E)$
(solid line) and $\rho_B(\br,E)$ (dashed line) vs $E/E_b$, for parameters
$h_0=1$\ nm, $b=8$\ nm ($E_b\simeq 0.082$\ eV), $g_v=7$\ eV, and $g_s=3$\ eV.
The LDOS for pristine graphene is included for reference (dotted lines).
Particle-hole symmetry is broken due to the presence of the scalar potential, as
seen most clearly in panels (i) and (j). The contribution from the vector
potential to the LDOS change obtained through Eq.\ \eqref{eq:G_tau:Born} is
proportional to $\sin\phi_d$ and therefore involves destructive interference
between scattering processes at every pair of locations symmetrically positioned
at equal perpendicular distances from the fold axis.
At $y=0$, equal scattering strengths lead to perfect cancellation and
$\Delta\rho_A^v(\br,E)=0$ for all $E$. For other points inside the
deformation region however, such cancellation does not occur. 

Crossings between $\Delta\rho_A(\br,E)$ and $\Delta\rho_B(\br,E)$ occur at
energies where $\Delta\rho_{\ell}^v(\br,E)$ (which oscillates due to the
interference processes mentioned above) passes through zero. For locations
further from the symmetry axis of the fold [see, for example, panels (k)
and (l)], some of the crossings are replaced by anticrossings.

\section{Kondo Physics}
\label{sec:Kondo}

In this section, we consider a magnetic adatom on a distorted graphene
membrane and find the effect of deformations on the characteristic Kondo
screening temperature $T_K$. We focus on the most basic models for top-
and hollow-site adsorption, disregarding additional complexities such as
orbital degrees of freedom and coupling anisotropies \cite{RMozara2018}
that have been shown to be important for certain adatoms. Our intention is to
emphasize general Kondo signatures, independent of the nature of the adsorbate
or the microscopic details of the hybridization. Section \ref{subsec:Kondo-model}
describes the Anderson impurity model used in our work and reviews certain
properties of similar models for impurities in a conventional metallic host.
Section \ref{subsec:top} applies the model to top-site adsorption on graphene,
while Sec.\ \ref{subsec:hollow} addresses the hollow-site case.

\subsection{Anderson impurity model}
\label{subsec:Kondo-model}

We investigate Kondo physics using a non-degenerate (single-orbital) Anderson
impurity Hamiltonian for a magnetic adatom hybridized with a strained graphene
host:
\begin{equation}
H = H_{\text{host}} + H_{\text{imp}} + H_{\text{hyb}}\,.
\label{eq:H_A}
\end{equation}
The host term is 
\begin{equation}
H_{\text{host}} = \sum_{\nu, \sigma} \veps_{\nu} \,
  c_{\nu, \sigma}^{\dag} c_{\nu, \sigma}^{\pdag}
\label{eq:H_host1}
\end{equation}
where $c_{\nu, \sigma}$ annihilates an electron in graphene with spin $z$
projection $\sigma = \pm 1/2$ (or equivalently, $\uparrow\!\!/\!\!\downarrow$),
non-spin quantum numbers that we collectively label $\nu$, and energy
$\veps_{\nu}$. The isolated adatom (or ``impurity'') is described by
\begin{equation}
H_{\text{imp}} = \veps_d \sum_{\sigma} d_{\sigma}^{\dag} d_{\sigma}^{\pdag}
  + U \, d_{\uparrow}^{\dag}  d_{\uparrow}^{\pdag} d_{\downarrow}^{\dag}
	  d_{\downarrow}^{\pdag} ,
\label{eq:H_imp}
\end{equation}
where $d_{\sigma}$ annihilates an electron of energy $\veps_d$ and spin $\sigma$
in an orbitally non-degenerate level having an on-site Coulomb repulsion $U$.
The coupling between the adatom and its host is captured in the term
\begin{equation}
H_{\mathrm{hyb}} = \sum_{j, \nu, \sigma} W_j d_{\sigma}^{\dag}
  \varphi_{\nu}(\bR_j) c_{\nu, \sigma}^{\pdag} + \text{H.c.} ,
\label{eq:H_hybrid1}
\end{equation} 
where $W_j$ is the matrix element (assumed to be spin-independent) for
tunneling into the impurity level from the $p_z$ orbital of carbon atom $j$
at location $\bR_j$ where the host eigenstate $\nu$ has a (dimensionless)
tight-binding wave function $\varphi_{\nu}(\bR_{j})$.
The $j$ sum runs over all carbon atoms in the graphene, even though $W_j$
will be non-negligible only for a small number of carbons located
close to the adatom. The wave function $\varphi_{\nu}$ is defined only at the
carbon sites, and is normalized so that 
\begin{equation}
\langle\varphi_{\nu}|\varphi_{\nu'}\rangle
  = \sum_j \varphi_{\nu}^*(\bR_j) \varphi_{\nu'}(\bR_j) = \delta_{\nu,\nu'}.
\end{equation}
In undeformed graphene, host eigenfunctions with small wave vectors
$\bq=(q\cos\phi_q,q\sin\phi_q)$ measured from the valley center at $\tau\bK$
($\tau=\pm 1$) and small energies $E=s_E\hbar v_F q$ ($s_E=\pm 1$) measured
from the Dirac point can be written in the form
\begin{equation}
\varphi_{\ell,\tau,\bq,s_E}(\bR_{j})
  = \frac{s_E^{\ell-1}}{\sqrt{N_c}} \, e^{i(\tau\bK+\bq)\cdot\bR_j} \,
	  e^{i\tau(\ell-3/2)\phi_q} ,
\label{eq:varphi:pristine}
\end{equation}
where $N_c$ is the number of unit cells in the graphene sheet, and
$\ell=1\, (A)$ or $2\, (B)$ labels the sublattice to which carbon atom $j$
belongs.

All single-particle energies ($\veps_{\nu}$, $\veps_d$, and the chemical
potential $\mu$) will be measured from the Dirac point in undistorted graphene.
We will focus on situations where $\veps_d-\mu<0$ and $2(\veps_d-\mu)+U>0$ so
that the ground state of the isolated impurity has a single electron that
therefore forms a local magnetic moment.

Equation \eqref{eq:H_hybrid1} can be rewritten as
\begin{equation}
H_{\mathrm{hyb}} = \frac{1}{\sqrt{N_c}} \sum_{\nu, \sigma} \tW_{\nu} \,
  d_{\sigma}^{\dag} c_{\nu, \sigma}^{\pdag} + \text{H.c.} ,
\label{eq:H_hybrid2}
\end{equation}
where
\begin{equation}
\tW_{\nu} = \sqrt{N_c} \sum_j W_j \, \varphi_{\nu}(\bR_j).
\label{eq:W_nu}
\end{equation}
It is convenient to transform to an energy representation by defining
\begin{equation}
a_{E, \sigma}^{\pdag} = \sqrt{\frac{\pi}{N_c g(E)}}
  \sum_{\nu} \delta(E - \veps_{\nu}) \, \tW_{\nu} c_{\nu, \sigma} ,
\end{equation}
with a (non-negative) hybridization function
\begin{equation}
g(E) = \frac{\pi}{N_c} \sum_{\nu} |\tW_{\nu}|^2 \,
  \delta(E - \veps_{\nu})
\label{eq:g:def}
\end{equation}
so that
$\{ a_{E, \sigma}^{\pdag}, \, a_{E',\sigma'}^{\dag} \}
  = \delta(E-E') \, \delta_{\sigma,\sigma'}$.
This allows one to express Eq.\ \eqref{eq:H_hybrid2} in the form
\begin{equation}
H_{\mathrm{hyb}} = \sum_{\sigma} d_{\sigma}^{\dag} \int \!\! dE \,
  \sqrt{g(E)/\pi} \, a_{E, \sigma}^{\pdag} + \text{H.c.} ,
\label{eq:H_hybrid3}
\end{equation}
and Eq.\ \eqref{eq:H_host1} in the form
\begin{equation}
H_{\text{host}} = \sum_{\sigma} \int \!\! dE \, E \, a_{E, \sigma}^{\dag}
  a_{E, \sigma}^{\pdag} + \ldots,
\label{eq:H_host2}
\end{equation}
where ``$\ldots$'' represents contributions from linear combinations of
host states that do not couple to the impurity and that will not be
considered any further.

Different possible symmetries of the impurity orbital and adsorption
configurations on the surface of graphene can be modeled by appropriate
choices of the matrix elements $W_j$ entering Eq.\ \eqref{eq:H_hybrid1}.
Both experiments and \textit{ab-initio} calculations suggest that two
adsorption configurations are energetically most likely \cite{YVirgus2014}:

\noindent
(1) ``Top-site'' attachment over a single carbon atom, has been observed for
Co adatoms on epitaxial monolayer graphene on SiC(0001), as well as
for both Co and Ni on quasi-freestanding mononolayer graphene on
SiC(0001) \cite{TEelbo20132}.
This configuration can be minimally described by just one nonzero $W_j$.

\noindent
(2) ``Hollow-site'' attachment at the center of a carbon hexagon, as has been
observed for Ni adatoms on monolayer graphene on SiC(0001) \cite{Gyamfi2012,%
TEelbo20132} and (in addition to top-site attachment) for both Co and Ni on
quasi-freestanding monolayer graphene on SiC(0001) \cite{TEelbo20132}.
This case can be approximated by six nonzero values $W_j$, which may all be
equal (e.g., for an $s$ or $d_{zz}$ impurity orbital) or may differ (as
in the case of other $d$ orbitals or any $f$ orbital).

Anderson models for these two adsorption configurations on undistorted
graphene have been considered previously; see, for example, Refs.\
\onlinecite{Uchoa2009,Zhu2010,Uchoa2011,Lo2014,Ruiz-Tijerina2016,%
Ruiz-Tijerina2017}. In this paper we generalize these previous treatments
to take into account deformation of the host surface.

Before discussing specific adsorption configurations, some general remarks
are in order. Equations \eqref{eq:H_A}, \eqref{eq:H_imp}, \eqref{eq:H_hybrid3},
and \eqref{eq:H_host2} together make up a standard representation of the
Anderson impurity model for a magnetic impurity hybridizing with a host via
an energy-dependent hybridization function $g(E)$. The canonical version
of this model has chemical potential $\mu=0$ and a ```top-hat'' hybridization
function
\begin{equation}
g(E) = \Gamma\,\Theta(D-|E|),
\label{eq:g:flat}
\end{equation}
where $\Theta(x)$ is the Heaviside function and the prefactor $\Gamma$ is termed
the ``hybridization width.'' For any $\Gamma>0$, the impurity spin degree of
freedom becomes collectively screened by the conduction band at temperatures
$T$ below a crossover scale: the Kondo temperature $T_K$. The dependences of
physical properties on $T$, magnetic field $B$, and frequency $\omega$ are
described by universal functions of $T/T_K$, $B/T_K$, and $\omega/T_K$ for
$T,B,\omega\lesssim T_K$
\footnote{We work in units where $\hbar = k_B = \mu_B = 1$}.
Provided that $U\ll D$, $\veps_d-\mu\ll-\Gamma$, $U+\veps_d-\mu\gg\Gamma$,
and $2U\Gamma\ll\pi|\veps_d-\mu|(U+\veps_d-\mu)$ (conditions that place
the model deep in its strongly correlated Kondo regime), the Kondo temperature
can be written \cite{Haldane1978}
\begin{equation}
T_K \simeq 0.36 \sqrt{\frac{2 U g(\mu)}{\pi}}
  \exp\Biggl[-\frac{\pi |\veps_d\!-\!\mu| (U\!+\!\veps_d\!-\!\mu)}
	                 {2Ug(\mu)}\Biggr] ,
\label{eq:T_K:Haldane}
\end{equation}
where $\mu=0$ and $g(\mu)=\Gamma$ in the canonical version of the model.
In more general cases where $\mu\ne 0$ and/or $g(E)$ is not strictly
constant---but still varies slowly within the energy range $|E-\mu|\lesssim U$
that sets $T_K$ \cite{Haldane1978}---the low-energy properties still follow
the universal scaling forms with a Kondo scale given (up to an overall
multiplicative correction) by Eq.\ \eqref{eq:T_K:Haldane}.

Anderson models in which $g(E)$ has strong energy dependence near the
chemical potential can exhibit strong deviations from canonical Kondo
physics. A well-studied example is the pseudogap Anderson model
\cite{DWithoff1990,Bulla1997,CGonzalez-Buxton1998,PCornaglia2009,Uchoa2009,
Li2013,LFritz2013,JJobst2013}, characterized by a hybridization function
\begin{equation}
g(E) = \Gamma \, |E/D|^r \, \Theta(D-|E|)
\label{eq:g:power-law}
\end{equation}
with a band exponent $r>0$.
If the chemical potential $\mu$ is nonzero so that $g(\mu)>0$, then the
pseudogap Anderson model exhibits conventional physics for
$T,\, B,\, |\omega| \lesssim T_K$, in many cases also retaining an
exponential dependence of $T_K$ on $g(\mu)$. For $\mu=0$, by contrast, the
depletion of hybridization close to the chemical potential allows Kondo
screening of the impurity moment only if the hybridization width exceeds a
threshold value $\Gamma_c>0$. For $\Gamma < \Gamma_c$, $T_K$ effectively
vanishes and the system instead approaches a low-energy regime in which the
impurity moment asymptotically decouples from the conduction band. A quantum
phase transition (QPT) at $\Gamma=\Gamma_c$ separates local-moment
($\Gamma<\Gamma_c$) and Kondo ($\Gamma>\Gamma_c$) phases. In each phase,
physical properties take scaling forms that depend on the band exponent $r$
entering Eq.\ \eqref{eq:g:power-law} as well as $T/T^*$, $B/T^*$, and
$\omega/T^*$. Here, $T^*$ (which replaces $T_K$ in the conventional Anderson
model) is a many-body scale that vanishes as
$T^* \propto |\Gamma-\Gamma_c|^{\nu}$ close to the QPT, with $\nu$ being a
positive, $r$-dependent exponent.

As pointed out previously \cite{Uchoa2009,Uchoa2011,Lo2014,Ruiz-Tijerina2016,
Ruiz-Tijerina2017}, adsorption of a magnetic impurity in a top-site
configuration on undeformed graphene can be described by an Anderson model
with a hybridization function that at low energies $|E|\ll D=3t_0$ corresponds
to Eq.\ \eqref{eq:g:power-law} with $r=1$, while hollow-site adsorption
realizes the case $r=3$. This raises the prospect of realizing the pseudogap
Kondo effect in undoped graphene where the chemical potential coincides with
the Dirac points, but (as mentioned in Sec.\ \ref{sec:intro}) there is a high
likelihood that the characteristic scale $T^*$ lies below the range accessible
in experiments. Our focus in this work is on a different regime $|\mu|=O(E_b)$
that reveals unique features of Kondo physics in deformed graphene. Here, the
low-energy properties follow their conventional (metallic) forms and the
effects of strain can be captured in the variation of the Kondo temperature
$T_K$ for different adatom locations relative to the peak deformation.

To calculate $T_K$, we solve the appropriate Anderson impurity model using
the numerical renormalization-group (NRG) method \cite{KWilson1975,
Krishna-murthy1980I,Krishna-murthy1980II,RBulla2008}, as adapted to treat an
arbitrary hybridization function \cite{CGonzalez-Buxton1998}. The Kondo
temperature is determined via the standard operational definition (with
$g\mu_B=k_B=1$) $T_K \chi_{\text{imp}}(T_K) = 0.0701$ \cite{KWilson1975,
Krishna-murthy1980I}, where $\chi_{\text{imp}}(T)$ is the impurity contribution
to the system's magnetic susceptibility at absolute temperature $T$. All NRG
results reported below were obtained using a Wilson discretization parameter
$\Lambda = 2.5$, retaining up to $N_{\mathrm{kept}}=2\ 000$ many-body
spin-multiplets after each iteration. A known artifact of NRG band
discretization is a reduction in the hybridization width from its nominal
value $\Gamma$ to an effective one $\Gamma/A$ \cite{Krishna-murthy1980I}.
When making comparisons with Eq.\ \eqref{eq:T_K:Haldane}, we use the correction
factor $A=1.204$ \cite{CGonzalez-Buxton1998} appropriate for $\Lambda = 2.5$
and a linear hybridization function [Eq.\ \eqref{eq:g:power-law} with $r=1$].

\subsection{Top-site adsorption}
\label{subsec:top}

When a magnetic atom adsorbs directly over a carbon atom, it is a good
approximation to assume that there is just one non-negligible hybridization
matrix element $W$. If the hybridizing carbon atom is at position $\bR$ in
sublattice $\ell$, then the relevant Anderson impurity model has hybridization
function
\begin{equation}
g_{\ts}(E)= 2 D \Gamma \, A_c \rho_{\ell}(\bR,E),
\label{eq:g:top}
\end{equation}
where $A_c=3\sqrt{3} a^2/2$ is the graphene unit cell area, $D = 3 t_0$ is the
half-bandwidth of graphene, $\Gamma=\pi W^2/(2D)$ is the mean value of
$g_{\ts}(E)$ taken over all $|E|<D$, and $\rho_{\ell}(\bR,E)$ is the
valley-summed local density of states per spin orientation, per unit area as
discussed in Secs.\ \ref{subsec:LDOS-bubble} and \ref{subsec:LDOS-fold}.
Equation \eqref{eq:g:top} shows that the hybridization function $g_{\ts}(E)$
directly follows the energy dependence of the LDOS for the sublattice to which
the hybridizing carbon belongs. In undeformed graphene, $\rho_{\ell}(\bR,E)$
reduces for $|E|\ll D$ to $\rho_0(E)$ given in Eq.\ \eqref{eq:rho_0:lin}, and
thus, $g_{\ts}(E)=(6\sqrt{3}/\pi) \Gamma |E/D|$.
We assume that the slowly varying out-of-plane deformations considered
in this study induce negligible change in the hybridization matrix element $W$,
so that strains enter the Anderson model solely through changes in
$\rho_{\ell}(\bR,E)$.

In Secs.\ \ref{subsubsec:top:bubble} and \ref{subsubsec:top:fold} below, we
present results for mechanical deformations with the same geometric parameters
as were used in Sec.\ \ref{sec:LDOS}: maximum height $h_0 = 1$\ nm, Gaussian
halfwidth $b = 8$\ nm (so that $E_b = \hbar v_F / b = 0.082$\ eV), and
gauge couplings $g_v=7$\ eV and $g_s=3$\ eV.  In light of the disagreement in
the literature over the sign and magnitude of $g_s$, we also show results for
$g_s = -3$\ eV and $g_s = 1$\ eV. We consider situations where the graphene is
gated or doped to produce a chemical potential
$\mu=\pm 0.15$\ eV $\simeq \pm 1.8 E_b$ in the energy range of largest
deformation-induced changes in the LDOS $\rho_{\ell}(\bR,E)$ (see Secs.\
\ref{subsec:LDOS-bubble} and \ref{subsec:LDOS-fold}).

It is also necessary to choose parameters $\veps_d$, $U$, and $\Gamma$
describing the adatom. To determine the parameter values appropriate for a
particular magnetic impurity species would require \textit{ab-initio}
calculations or detailed experimental measurements that are beyond the scope of
this work.
However, qualitative behaviors to be expected can be adequately illustrated by
focusing on a single value of the level energy $\veps_d = -1$\ eV with either
$U = -2\veps_d = 2$\ eV (for an impurity level that is particle-hole-symmetric
for $\mu=0$) or $U=\infty$ (representing maximal particle-hole asymmetry). We
choose $0.65\ \text{eV}\le\Gamma\le3\ \text{eV}$, values that cause the Kondo
temperature $T_K^0$ in the absence of strain to fall between 20\ mK and 4.2\ K.

Let us start from the reference case of an adatom with parameters
$U=-2\veps_d=\Gamma=2$\ eV adsorbed on top of a carbon atom in undeformed
graphene having a chemical potential $\mu=\pm 0.15$\ eV. NRG calculations for
this case give $T_K^0=0.21$\ K, within 20\% of the value 0.25\ K predicted by
Eq.\ \eqref{eq:T_K:Haldane}. This close agreement suggests that, despite the
complicated energy dependence of $g_{\ts}(E)$, the Kondo scale is set mainly
by the value of $g_{\ts}(\mu)$. (We will return to this point when we discuss
hollow-site adsorption.) That $T_K^0$ is independent of the sign of $\mu$ is
due to the strict particle-hole symmetry about the Dirac points $\veps=0$ of
the hybridization function [i.e., $g_{\ts}(E) = g_{\ts}(-E)$ for all $E$] and
of the adatom energy levels (i.e., $U=-2\veps_d$).

Having established this reference case, we can now look at the effects of
deformation of the graphene host.

\subsubsection{Kondo temperature for top-site adsorption near a Gaussian bubble}
\label{subsubsec:top:bubble}

\begin{figure}[t]
\includegraphics[width=3.35in]{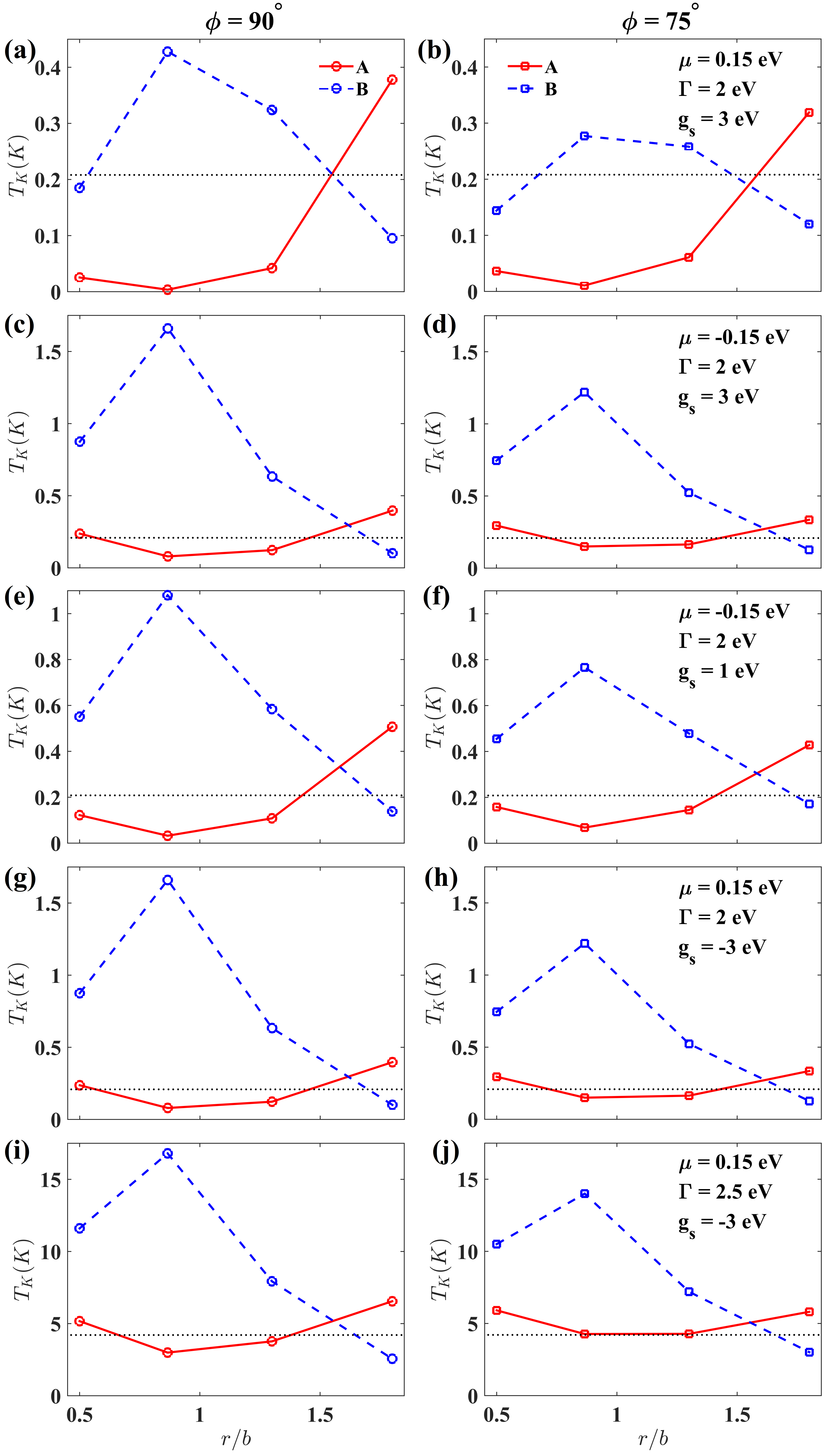} 
\caption{\label{Fig4}%
Kondo temperature $T_K$ vs distance $r$ from the center of a Gaussian
bubble deformation [Fig.\ \ref{Fig1}(a)] along directions $\phi=90^{\circ}$
(left panels) and $\phi=75^{\circ}$ (right panels). Data are for height
$h_0=1$\ nm and width $b=8$\ nm, for gauge vector coupling $g_v = 7$\ eV, and
for a symmetric magnetic impurity described by $U = -2\veps_d = 2$\ eV.
The other model parameters are specified in a legend for each row. Dashed (solid)
lines connect $T_K$ values for adatoms on top of carbon atoms in the $A$ ($B$)
sublattice. A horizontal dotted line represents the Kondo temperature in the
absence of deformation.
}
\end{figure}

Figure \ref{Fig4} shows Kondo temperatures for an adatom on top of a carbon atom
in sublattice $A$ (data points connected by solid lines) or in sublattice $B$
(dashed lines), located at four distances $r$ from the peak of a Gaussian bubble
along directions at $\phi=90^{\circ}$ (left panels) and $\phi=75^{\circ}$ (right
panels) measured counterclockwise from the positive $x$ axis. The locations
illustrated are marked by white dots in Fig.\ \ref{Fig1}(c) and correspond to
the ones in Fig.\ \ref{Fig2}. All data in this figure are
for $U=-2\veps_d=2$\ eV. Each row corresponds to a different combination of the
chemical potential $\mu$, the adatom hybridization width $\Gamma$, and the
scalar gauge coupling $g_s$. Each row after the first differs from a higher
row by a change in just one of $\mu$, $\Gamma$, and $g_s$, chosen to illustrate
and highlight a general trend as one moves within the parameter space of the
model.

Figs.\ \ref{Fig4}(a) and \ref{Fig4}(b) illustrate the case $g_s = 3$\ eV
for which the $A$- and $B$-sublattice LDOS are plotted in Fig.\ \ref{Fig2}.
The other parameters are $\mu=0.15\ \text{eV} \simeq 1.8 E_b$ and
$\Gamma=2$\ eV. At each of the eight locations illustrated, the
scalar gauge potential decreases the LDOS on both sublattices, while the
vector gauge potential increases the LDOS on one sublattice and reduces it
on the other sublattice, as expected from Eqs.\ \eqref{eq:Drho_s:symm} and
\eqref{eq:Drho_v:symm}; since the vector gauge coupling is larger than the
scalar, the net effect at all locations except $(r,\phi)=(0.5b,75^{\circ})$
is a net increase of $\rho_{\ell}(\bR,\mu)$ for one sublattice and a net
decrease for the other. Comparison between Figs.\ \ref{Fig2}(e)--\ref{Fig2}(l),
\ref{Fig4}(a), and \ref{Fig4}(b) reveals that $T_K$ rises/falls in close
correlation with the value of the sublattice LDOS at the chemical potential.
The extremal $T_K$ values occur at $r=0.866b$, $\phi=90^{\circ}$, where the
$A$ sublattice has a density of states per unit area $\rho_A(\mu)=0.0176/(DA_c)$
and a Kondo scale $T_K = 3.5$\ mK, while the $B$ sublattice has
$\rho_B(\mu)=0.0312/(DA_c)$ and $T_K = 0.43$\ K; for comparison, pristine
graphene has $\rho_0(\mu)=0.0268/(DA_c)$ and (as noted above) Kondo temperature
$T_K^0=0.21$\ K. In this particular region near a bubble deformation, the LDOS
at the chemical potential decreases by 34\% on the $A$ sublattice and increases
by 16\% on the $B$ sublattice. These changes are amplified in the Kondo
temperature, which (relative to undeformed graphene) decreases by a factor of
60 on the $A$ sublattice while doubling on the $B$ sublattice. However, the
amplifications are not quite as strong as the 110-fold decrease for $A$ and the
3.7-fold increase for $B$ predicted by Eq.\ \eqref{eq:T_K:Haldane}, reinforcing
the point that $T_K$ depends on values taken by the hybridization function
$g_{\ts}(E)$ within a window around the chemical potential, not just on
$g_{\ts}(\mu)$. Note in particular that, as can be seen in Fig.\ \ref{Fig2} and
Eq.\ \eqref{eq:g:top}, a deformation that decreases (increases) $g_{\ts}(\mu)$
tends to increase (decrease) $g_{\ts}(E)$ at energies $E$ not too far from
$\mu$. The subtle interplay of these changes in $g_{\ts}(E)$ explains,
for example, why deformation results in a modest decrease of $T_K$ on the
$B$ sublattice at $r=0.5b$, $\phi=90^{\circ}$ even though $\rho_B(\br,\mu)$
and hence $g_{\ts}(\mu)$ undergo a slight increase.

Panels (c) and (d) in Fig.\ \ref{Fig4} illustrate the same situation as
panels (a) and (b), respectively, apart from a reversal in sign of the
chemical potential to $\mu = -0.15$\ eV. As noted above, the Kondo
temperature for our reference case in undeformed graphene is unchanged
by this reversal due to the strict particle-hole symmetry of the LDOS
and the adatom level energies. However, the scalar component of the LDOS
changes induced by deformation breaks particle-hole symmetry; for $g_s>0$,
the effect is to decrease the LDOS for $E>0$ and increase it for $E<0$.
Therefore, the case $\mu=-0.15$\ eV samples a higher LDOS in the vicinity
of the chemical potential than is the case for $\mu=0.15$\ eV, and as one
might expect, higher Kondo temperatures follow. The highest and lowest
Kondo temperatures in panels (c) and (d) are 1.7\ K and 79\ mK,
respectively 8 and 0.4 times $T_K^0$. That a modest (here $0.6\%$) strain
can enhance $T_K$ by an order of magnitude is one of the principal findings
of this work. It significantly improves the prospects of experimental
detection of Kondo physics in situations where the signatures would
otherwise occur below the base temperature of an experiment.

Panels (e) and (f) in Fig.\ \ref{Fig4} differ from panels (c) and (d) only
by an decrease in $g_s$ from 3\ eV to 1\ eV, which reduces the magnitude of
the particle-hole symmetry-breaking caused by the scalar potential.
The variation of $T_K$ with position $(r,\phi)$ in panels (e) and (f) is
qualitatively very similar to that in panels (c) and (d). However, each
$T_K$ in the third row of the figure is smaller than its counterpart
in the second row, while still being greater than the corresponding value
for $\mu=0.15$\ eV in the first row.

Panels (g) and (h) in Fig.\ \ref{Fig4} differ from panels (a) and (b) only
by a switch in $g_s$ from $3$\ eV to $-3$\ eV, reversing the sign of the
LDOS change due to the scalar potential while leaving unaffected the
change due to the vector potential. For the cases considered here, where
the undeformed $g_{\ts}(E)=g_{\ts}(-E)$ and $U=-2\veps_d$, a change in sign
of $g_s$ while keeping $\mu$ constant has the same effect on $T_K$ as a change
in the sign of $\mu$ at fixed $g_s$. For this reason, the Kondo temperatures
shown in panels (g) and (h) are identical to those in panels (c) and (d).

Finally in Fig.\ \ref{Fig4}, panels (i) and (j) differ from panels (g) and (h)
only by an increase in $\Gamma$ from 2\ eV to 2.5 \ eV. This change increases
the Kondo temperature $T_K^0$ for an undeformed host from $0.21$\ K to $4.2$\ K.
Near the Gaussian bubble, the pattern of $T_K$ values on each sublattice is
qualitatively very similar to that for $\Gamma=2$\ eV. However, panels (i) and
(j) show values of $T_K/T_K^0$ spanning a range 0.61 to 4.0 that is narrower than
the range 0.38 to 8.0 in panels (g) and (h). Such a reduction with increasing
$\Gamma$ in the sensitivity of the Kondo scale to deformation-induced LDOS
changes is consistent with the approximation that $T_K$ is given by
Eqs.\ \eqref{eq:T_K:Haldane} and \eqref{eq:g:top}. For still greater values of the
hybridization width, the system should cross over from its Kondo regime into mixed
valence, where $T_K$ depends linearly---rather than exponentially---on $g(\mu)$.

\subsubsection{Kondo temperature for top-site adsorption near a long Gaussian fold}
\label{subsubsec:top:fold}

\begin{figure}[t]
\includegraphics[width=3.38in]{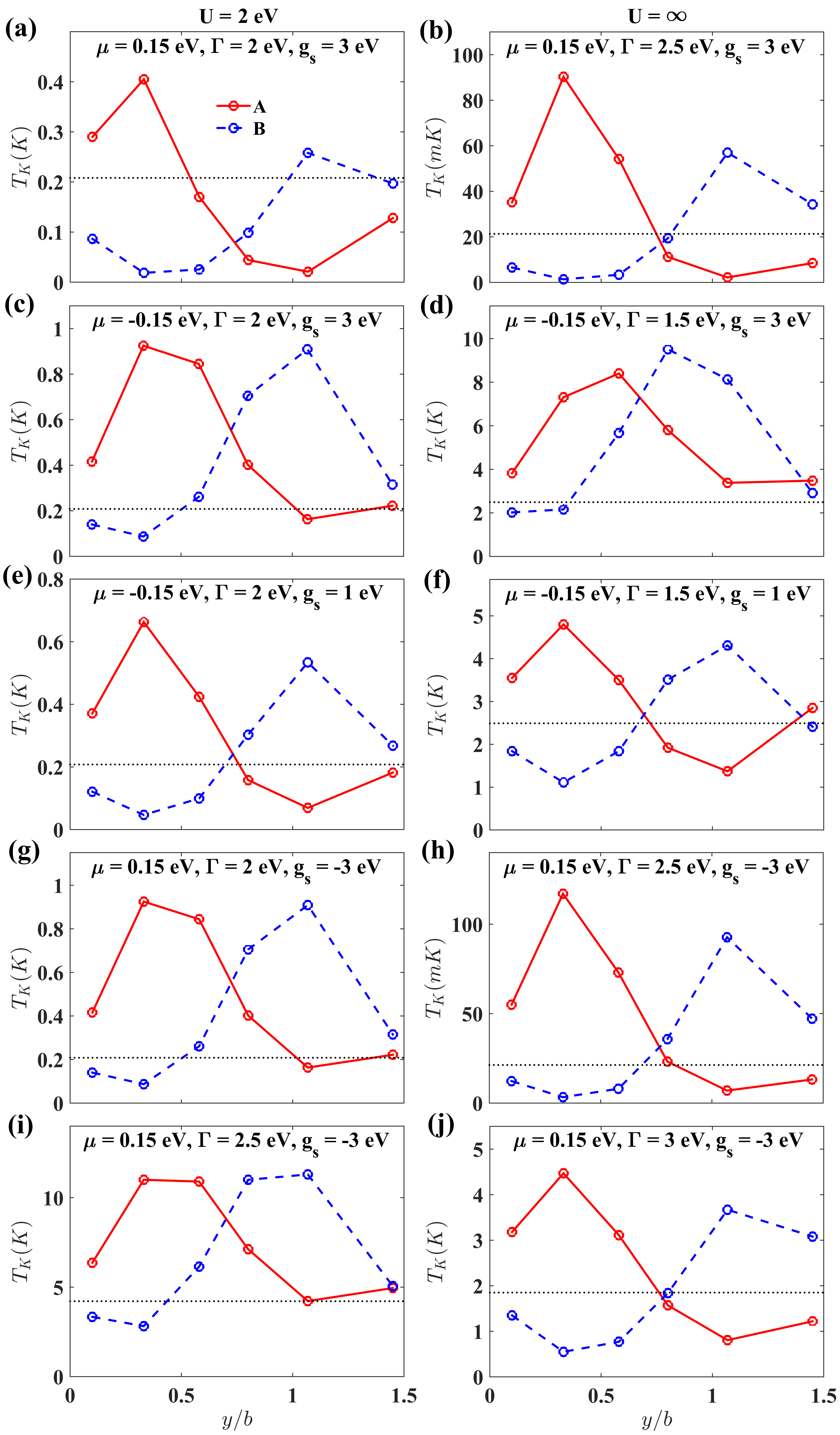} 
\caption{\label{Fig5}%
Kondo temperature $T_K$ vs distance $y$ away from the symmetry axis of an
extended Gaussian fold deformation [Fig.\ \ref{Fig1}(b)]. Data are for
height $h_0=1$\ nm and width $b=8$\ nm, for gauge vector coupling
$g_v = 7$\ eV, for an impurity level energy $\veps_d = -1$\ eV,
and for on-site Coulomb interactions $U=2$\ eV (left panels) and $U=\infty$
(right panels). All other model parameters are specified in a legend for
each row. Those for the left panels are identical to the ones in the
corresponding row of Fig.\ \ref{Fig4}. Each panel in the right column shares
the same $\mu$ and $g_s$ as its left neighbor, but has a different $\Gamma$.
Dashed (solid) lines connect $T_K$ values for adatoms on top of carbon atoms
in the $A$ ($B$) sublattice, and a horizontal dotted line represents the Kondo
temperature in the absence of deformation.
}
\end{figure}

We now turn to Fig.\ \ref{Fig5}, which shows Kondo temperatures for an adatom
on top of a carbon atom in sublattice $A$ (solid lines) or sublattice $B$
(dashed lines) at six perpendicular distances $y$ from the symmetry axis ($y=0$)
of an extended Gaussian fold. The locations illustrated are marked by white
dots in Fig.\ \ref{Fig1}(d) and are the ones for which the LDOS is plotted in
Fig.\ \ref{Fig3}. The left panel in each row shows data for the same combination
of $\mu$, $g_s$, $U=-2\veps_d$, and $\Gamma$ (and hence the same value of
$T_K^0$) as appears in the corresponding row of Fig.\ \ref{Fig4}, thereby
facilitating the identification of similarities and differences
between the effects of bubble and fold deformations. The right panel differs
from its left counterpart only in that the data are for $U=\infty$ and $\Gamma$
has been adjusted to keep $T_K^0$, the Kondo temperature in the absence of
deformation, within the range 20\ mK to 2--3\ K. (Without any adjustment
of $\Gamma$, the value of $T_K^0$ calculated for $U=\infty$ and our standard
hybridization width $\Gamma=2$\ eV would be 24\ $\mu$K for chemical potential
$\mu=0.15$\ eV or 490\ K for $\mu=-0.15$\ eV, in both cases placing the
Kondo scale outside the window of experimental interest for most experiments.)

Figs.\ \ref{Fig5}(a)--\ref{Fig5}(d) illustrate the case $g_s = 3$\ eV
for which the $A$- and $B$-sublattice LDOS are plotted in Fig.\ \ref{Fig3}.
Panels (a) and (b) are for $\mu=0.15\ \text{eV} \simeq 1.8 E_b$, while
(c) and (d) are for $\mu=-0.15$\ eV. Comparison between these panels and
Figs.\ \ref{Fig3}(g)--\ref{Fig3}(l) reveals that (just as for the Gaussian
bubble illustrated in Figs.\ \ref{Fig2} and \ref{Fig4}), $T_K$ varies within
a panel in close correlation with the value of $\rho_{\ell}(\bR,\mu)$. Due to
a reversal in the sign of the vector gauge field compared to the bubble, for
$|y|\lesssim 0.7b$ it is sublattice $A$ (rather than $B$) that has the larger
LDOS and hence the higher Kondo temperature. Nonetheless, the range of values
of $T_K/T_K^0$ for $U=-2\veps_d$ shown in (a) and (c) is similar to that for
the bubble in Figs.\ \ref{Fig4}(a)--\ref{Fig4}(d). For $U=\infty$, there is
a much stronger difference between the behavior for $\mu=0.15$\ eV and
$\mu=-0.15$\ eV: in the former case, even with $\Gamma$ increased to 2.5\ eV
the Kondo temperature on undeformed graphene is only $T_K^0 = 21$\ mK, an order
of magnitude smaller than in our reference case $U=-2\veps_d=\Gamma=2$\ eV; for
$\mu=-0.15$\ eV, by contrast, even with $\Gamma$ reduced to 1.5\ eV, we
find an order-of-magnitude enhancement of $T_K^0$ to 2.5\ K. 

The asymmetric behavior seen for $U=\infty$ under reflection of the chemical
potential about the Dirac point can be understood within a poor man's scaling
analysis of the Anderson model \cite{Hewson1997,Cheng2017} that progressively
integrates out the conduction-band states lying furthest in energy from the
chemical potential, accounting for the eliminated band-edge states through
perturbative adjustment of $\veps_d$, $U$, and $g(E)$. For $U=\infty$,
the renormalized value of the level energy $\tveps_d$ evolves according to the
differential equation \cite{Cheng2017}
\begin{equation}
\label{eq:Ed:scaling}
\pi \frac{d\tveps_d}{d\tD}
  = \frac{\tG(\tD+\mu)}{\tD-(\tveps_d-\mu)}
  - \frac{2\tG(-\tD+\mu)}{\tD+(\tveps_d-\mu)} ,
\end{equation}
where $\tD$ (with initial value $D+|\mu|$ and satisfying $d\tD<0$) is the
running half-bandwidth as measured from the chemical potential and
$\tG(\pm\tD+\mu)$ represents the renormalized hybridization function evaluated
at the edges of the reduced band. The factor of 2 in the second term on the
right-hand side of Eq.\ \eqref{eq:Ed:scaling} arises because an electron of
either spin $z$ projection $s=\pm\frac{1}{2}$ can undergo virtual tunneling
from the bottom of the band into the empty impurity level. Virtual tunneling
of the electron from a singly occupied impurity level to the upper band edge
(described by the first term on the right side) has no factor of 2 because it
must conserve the spin of that electron. Equation \eqref{eq:Ed:scaling} shows
that band states below, but not very far from, the chemical potential make a
greater contribution to the renormalization of $\tveps_d$ than do band states
an equal distance above the chemical potential. Due to the presence of the
Dirac point at $E=0$, the LDOS at energy $E=\mu-\tD$ for a given $\tD>0$ is
lower for $\mu=0.15$\ eV than it is for $\mu=-0.15$\ eV. A faster scaling of
$\tveps_d$ with decreasing $\tD$ generally results in a higher Kondo
temperature \cite{Cheng2017}, so it is to be expected that $T_K^0$ is higher
for $\mu<0$.

Panels (e) and (f) in Fig.\ \ref{Fig5} differ from panels (c) and (d) only by
an decrease in $g_s$ from 3\ eV to 1\ eV, which reduces the particle-hole
symmetry-breaking caused by the scalar potential. As was the case for
Fig.\ \ref{Fig4}, the variation of $T_K$ with position in panels (e) and (f)
is qualitatively similar to that in panels (a)--(d), but at a given location,
$T_K$ for each sublattice lies between the corresponding values in the first
and second rows of the figure.

Panel (g) in Fig.\ \ref{Fig5} differs from panel (a) only by a switch in the
sign of $g_s$. As discussed in connection with Figs.\ \ref{Fig4}(g) and
\ref{Fig4}(h), for $U=-2\veps_d$ this switch is equivalent to changing the sign
of $\mu$, implying that the data in Fig.\ \ref{Fig5}(g) are identical to those
in Fig.\ \ref{Fig5}(c). By contrast, there is no simple relation between the
$T_K$ values in panels (a) and (h), which differ not only as to the sign of
$g_s$, but also in their values of $\Gamma$. Even if the $\Gamma$ values were
the same, there would be no symmetry connecting these two $U=\infty$ cases.

Finally in Fig.\ \ref{Fig5}, panels (i) and (j) differ from panels (g) and (h),
respectively, only by an increase in $\Gamma$ by 0.5\ eV. This change increases
the Kondo temperature for an undeformed host from $T_K^0=0.21$\ K in (g) to
$T_K^0=4.2$\ K in (i), and from $T_K^0=21$\ mK in (j) to $T_K^0=1.9$\ K in
(j). The pattern of $T_K$ values on each sublattice is qualitatively very
similar for the smaller and larger $\Gamma$ values, but (just as is seen for
the bubble), the larger $\Gamma$ yields ratios $T_K/T_K^0$ that deviate less
from 1, indicating that as the system moves from deep in its Kondo regime
toward mixed valence, the Kondo temperature becomes less sensitive to
deformation-induced changes in the LDOS on each sublattice.

To summarize, Figs. \ref{Fig4} and \ref{Fig5} show many similarities
between the spatial variation of $T_K$ for top-site adsorption near a Gaussian
deformation of bubble and fold geometry. The most striking feature is that the
Kondo temperature, a quantity that can be deduced from scanning-tunneling
spectroscopy performed over adatom, serves to amplify deformation-induced
changes in the LDOS. The degree of enhancement or suppression of $T_K$ relative
to its value $T_K^0$ for undeformed graphene depends on properties of the
adatom (as modeled via the parameters $\veps_d$, $U$, and $\Gamma$), on
characteristics of the graphene (such as the parameters $g_s$ and $g_v$ and the
degree of strain), and on environmental details such as the chemical potential
$\mu$ established via doping or back-gating. However, without any fine-tuning of
parameters, we have demonstrated that $T_K$ can easily be enhanced by an
order of magnitude, increasing the prospects for experimental observation of
Kondo phenomenology.

In the top-site configuration, the effective scalar potential $\Phi$ defined in
Eq.\ \eqref{eq:Phi} tends to modify the Kondo temperature on both
sublattices in the same direction, lowering $T_K$ in situations where the
chemical potential has the same sign as $g_s$ but raising it when $\mu$ and
$g_s$ have the opposite sign. By contrast, the vector potential defined in Eq.\
\eqref{eq:A_x,y} changes $T_K$ in the opposite direction for adatoms attached
to the $A$ and $B$ sublattices, but the direction of change for a given
sublattice is unaffected by reversal in the sign of $\mu$. These differing
trends provide a signature that can unambiguously distinguish Kondo physics from
other phenomena that may occur in graphene.

On a more speculative level, the results in Figs. \ref{Fig4} and \ref{Fig5}
also suggest a possible method for disentangling the scalar and vector
contributions to the  deformation-induced LDOS change. Suppose that for a
given adatom species and a fixed chemical potential $\mu$, it is possible to
measure (e.g., via the width of an STM Fano lineshape) not only the Kondo
temperature $T_K^0$ for top-site adsorption on pristine graphene, but also the
scales $T_K^A(\bR)$ and $T_K^B(\bR)$ for adsorption above close-lying $A$- and
$B$-sublattice carbon atoms in the vicinity of a smooth deformation. As noted
above, and further discussed below in connection with Fig.\ \ref{Fig7},
Eq.\ \eqref{eq:T_K:Haldane} proves to be reasonably accurate for top-site
adsorption deep in the Kondo regime. Using Eq.\ \eqref{eq:g:top}, and
decomposing the LDOS in the presence of deformation according to Eq.\
\eqref{eq:LDOS:full}, one can estimate the fractional LDOS change due to the
scalar effective potential,
\begin{equation}
\Delta\rho_A^s(\bR) / \rho_0(\mu) \simeq c \ln[T_K^A(\bR) T_K^B(\bR)/(T_K^0)^2],
\label{eq:T_K:AB-prod}
\end{equation}
as well as its counterpart due to the vector effective potential,
\begin{equation}
\Delta\rho_A^v(\bR,\mu) / \rho_0(\mu) \simeq c \ln[T_K^A(\bR) / T_K^B(\bR)],
\label{eq:T_K:AB-ratio}
\end{equation}
where $c$ is a dimensionless constant that is independent of $\bR$. By
applying Eqs. \eqref{eq:T_K:AB-prod} and \eqref{eq:T_K:AB-ratio}  at different
points $\bR$ relative to the peak deformation, one should be able to gain
insight into the sign and magnitude of the coupling $g_s$ relative to the
better-understood quantity $g_v$.

Even though we have considered a relatively small number of combinations of
model parameters, the results in Figs. \ref{Fig4} and \ref{Fig5} are broadly
representative of the range of qualitative behaviors that can be expected
across the full parameter space. Additional complexities, such as adatoms
having higher spins or spin-anisotropic interactions, are likely to alter
only in quantitative detail the Kondo amplification of deformation-induced
LDOS changes that is the central result of this work.

\subsection{Hollow-site adsorption}
\label{subsec:hollow}

\begin{figure}[t]
\includegraphics[height=1.25in]{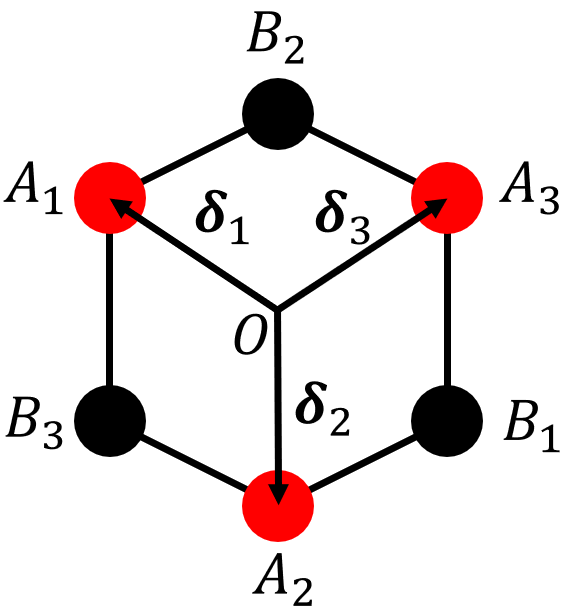}
\caption{\label{Fig6}%
Schematics of the hollow site geometry. The impurity adatom, located directly
above the point labeled $O$, hybridizes with the six nearest carbon atoms in the
plane represented as circles. Nearest carbon atoms $A_j$ ($j=1$, $2$, $3$) from
sublattice $A$ (red circles) are located at positions described by
two-dimensional vectors $\bdel_j$, while nearest carbon atoms $B_j$ from
sublattice $B$ (black circles) are located at $-\bdel_j$.
}
\end{figure}

The description of hollow-site adsorption is more complicated than that of the
top-site configuration due to quantum-mechanical interference between
tunneling from the active impurity level to different carbon atoms in
monolayer graphene. We first treat adsorption at a hollow site
on pristine graphene before considering the effects of deformation.

\subsubsection{Hollow-site adsorption on pristine graphene}
\label{subsubsec:hollow:pristine}

For simplicity, we assume that the adatom is located above point $\bR$ on
the graphene plane and hybridizes only with the six nearest carbon atoms
(see Fig.\ \ref{Fig6}): three from sublattice $A$ at locations
$\bR_{A,n}=\bR+\bdel_n$ ($n=1$, $2$, $3$) and three from sublattice $B$ at
locations $\bR_{B,n}=\bR-\bdel_n$, with $\bdel_1=\left(-\sqrt{3},1\right)a/2$,
$\bdel_2=\left(0,-1\right)a$, and $\bdel_3=\left(\sqrt{3},1\right)a/2$.
In general, the hybridization matrix elements $W_j$ between the active
impurity orbital and the nearest carbon atoms can take independent values
$W_{A,n}$ and $W_{B,n}$. In previous studies \cite{Uchoa2009,Uchoa2011,%
Ruiz-Tijerina2016,Ruiz-Tijerina2017}, the hybridization function
$g_{\hs}(E)$ for hollow-site adsorption was calculated by substituting
into Eq.\ \eqref{eq:g:def} the expression
\begin{equation}
\tW_{\nu}
  = \sum_{n=1}^{3} \left[ W_{A,n} \varphi_{A,\nu}(\bR_{A,n}) +
	  W_{B,n} \varphi_{B,\nu}(\bR_{B,n}) \right]
\label{eq:W_nu:hs}
\end{equation}
with $\nu=(\tau,\bq,s_E)$ and $\varphi_{\ell,\tau,\bq,s_E}(\br)$ being a
plane-wave state defined in Eq.\ \eqref{eq:varphi:pristine}.

Reference \onlinecite{Uchoa2011} identifies two different classes of
hollow-site adsorption. In situations where
\begin{equation}
W_{A,n}=W_A \quad \text{and} \quad
W_{B,n}=W_B \quad \text{for \;} n = 1, 2, 3,
\label{eq:C_3v}
\end{equation}
such that the adatom hybridizes equally with the three nearest carbons on a
given sublattice, $g_{\hs}(E) \propto |E/D|^3$ at low energies. In all other
cases, where the set of hybridization matrix elements breaks the full
$C_{3v}$ point-group symmetry of the lattice, one instead finds
$g_{\hs}(E) \propto |E/D|$, qualitatively the same as for top-site adsorption.
The additional factor of $(E/D)^2$ in the first class can be attributed to
the appearance in $|\tilde{W}_{\tau,\bq,s_E}|^2$ of a multiplicative factor
of $|\sum_{n=1}^3 \exp[i(\tau\bK+\bq)\cdot\bdel_n]|^2$, which vanishes at the
valley minimum $\bq=\mathbf{0}$.

Anticipating extension of the calculation to situations with deformation,
we can instead derive the hybridization function $g_{\hs}(E)$ from the
continuum-limit single-particle Green's function in the basis
$(\psi_A, \psi_B)$:
\begin{widetext}
\begin{equation}
G(\br,\br',E)
= \sum_{\nu,\nu'} \langle\br|\nu\rangle
   \langle\nu|(E + i s_E \,0^+ - H)^{-1}|\nu'\rangle
	 \langle \nu'|\br'\rangle
= \sum_{\nu} \frac{1} {E + i s_E 0^+ - \veps_{\nu}}
   \begin{pmatrix}
   \psi_{A,\nu}(\br)\psi_{A,\nu}^*(\br') & \psi_{A,\nu}\psi_{B,\nu}^*(\br')
	 \\[1ex]
   \psi_{B,\nu}(\br)\psi_{A,\nu}^*(\br') & \psi_{B,\nu}\psi_{B,\nu}^*(\br')
\end{pmatrix} .
\end{equation}
\end{widetext}
The continuum-limit wave functions $\psi_{\ell,\nu}(\br)$ are defined at all
two-dimensional position vectors $\br$ within the graphene sheet of total
area $A$, and are normalized so that
\begin{equation}
\langle\psi_{\ell,\nu}|\psi_{\ell,\nu'}\rangle
  = \int \!\! d^2\br \; \psi_{\nu}^* (\br) \, \psi_{\nu'}(\br)
	= \delta_{\nu,\nu'}.
\end{equation}
These continuum-limit wave functions can be connected with the tight-binding
ones entering Eq.\ \eqref{eq:W_nu} via $\psi_{\ell,\nu}(\br) =
A_c^{-1/2}\varphi_{\ell,\nu}(\br)$, where $A_c$ is the area of the graphene
unit cell. This allows one to write
\begin{equation}
g_{\hs}(E) = -s_E A_c \, \Im \!\!\!\! \sum_{n,n',\ell,\ell'} \!\!\!\!
  W_{\ell,n} W_{\ell'n'}^*
	\bigl[ G(\bR_{\ell,n}, \bR_{\ell',n'}, E)\bigr]_{\ell\ell'} .
\label{eq:g_hs}
\end{equation}
In the range $|E|\ll t_0$, one can reduce the above to a sum over
contributions from valley-resolved Green's functions $G_{\tau}(\br,\br',E)$
defined in Eq.\ \eqref{eq:G_tau:def}.

Equation \eqref{eq:g_hs} can be evaluated in closed form within the
approximation of linear dispersion about the Dirac points. Here, we illustrate
this by summarizing the results for configurations with full $C_{3v}$ symmetry
[i.e., satisfying Eq.\ \eqref{eq:C_3v}] where, moreover, $W_A/W_B$ is real.
This encompasses as special cases both (a) $W_A=W_B$ describing an active
impurity orbital that has cylindrical symmetry about an axis perpendicular to
the graphene plane, as is the case for $s$, $d_{zz}$, and $f_{z^3}$ orbitals,
and (b) $W_A=-W_B$, appropriate for $f_{x(x^2-3y^2)}$ and $f_{y(y^2-3x^2)}$
orbitals. After some laborious algebra, one finds that the $\ell\ne\ell'$
terms in Eq.\ \eqref{eq:g_hs} sum to zero, while the $\ell=\ell'$ terms
combine to give
\begin{equation}
g_{\hs}(E) = \frac{36\sqrt{3}}{\pi} \Gamma  |E/D|
  \bigl[ 1 - J_0\bigl(2\sqrt{3}|E|/D\bigr) \bigr]
\label{eq:g_hs:pristine}
\end{equation}
with $\Gamma = \pi(|W_A|^2 + |W_B|^2) / (4D)$. On the right-hand side of
Eq.\ \eqref{eq:g_hs:pristine}, the term 1 inside the square brackets comes
from pure LDOS ($n=n'$) terms in Eq.\ \eqref{eq:g_hs}, while the zeroth-order
Bessel function comes from nonlocal ($n\ne n'$) terms. In the regime where
$2\sqrt{3}|E|/D \ll 1$, the approximation $J_0(x)\simeq 1-x^2/4$ leads to
$g_{\hs}(E)\simeq(108\sqrt{3}/\pi) \Gamma |E/D|^3$, consistent with previous
work \cite{Uchoa2009,Uchoa2011,Ruiz-Tijerina2016,Ruiz-Tijerina2017}. The
additional factor of $(E/D)^2$ compared with $g_{\ts}(E)\propto|E/D|$ arises
from the complete destructive interference at the Dirac points ($E=0$)
between (i) virtual tunneling of an electron from a given carbon atom in
sublattice $\ell$ into the active impurity level, then from the impurity
back to the same carbon atom, and (ii) similar processes that end with the
electron tunneling back to one of the other two nearest carbons belonging to
sublattice $\ell$.

\begin{figure}[t]
\includegraphics[width=2.8in]{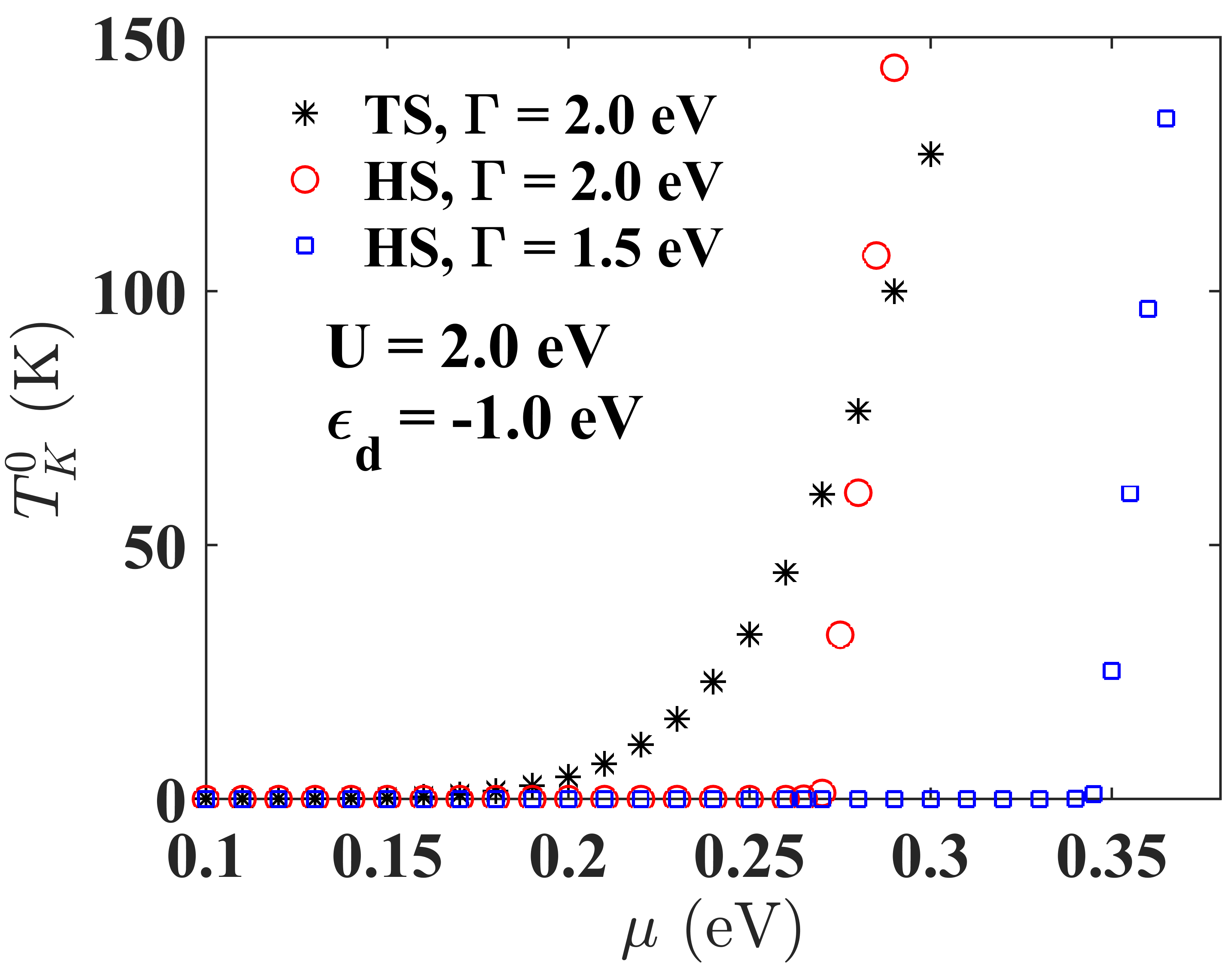}
\caption{\label{Fig7}%
Kondo temperature $T_K^0$ vs chemical potential $\mu$ for an adatom with
$U=-2\veps_d=2$\ eV on pristine (undeformed) graphene, showing data for
top-site adsorption with hybridization width $\Gamma = 2$\ eV
(\raisebox{2pt}{$\ast$}) as well as $C_{3v}$-symmetric hollow-site adsorption
with $\Gamma = 2$\ eV ({\large$\circ$}) and $\Gamma = 1.5$\ eV ($\square$).
}
\end{figure}

The cubic energy dependence of $g_{\hs}(E)$ for small $|E/D|$ means that
hybridization is greatly suppressed for hollow-site adsorption compared with
its top-site counterpart. Hence, for a chemical potential close to the Dirac
point and any given combination of the parameters $\veps_d$, $U$, and
$\Gamma$, the Kondo scale $T_K^0$ will generally be much lower in the
hollow-site configuration, as noted previously in Ref.\ \onlinecite{Lo2014}.
We illustrate this tendency in Fig.\ \ref{Fig7}, which plots the Kondo
temperature $T_K^0$ vs the chemical potential $\mu$ for three different
cases, all involving a magnetic impurity level with $U = -2\veps_d=2$\ eV.
Asterisks show $T_K^0$ for top-site adsorption of an impurity having
hybridization width $\Gamma=\pi W^2/(2D) = 0.2$\ eV. Over the range of
$\mu$ spanned in the figure, the numerical data agree with Eq.\
\eqref{eq:T_K:Haldane} to within better than a factor of $4$. This
observation supports the assessment made near the start of Sec.\
\ref{subsec:top} that the Kondo physics for top-site adsorption on graphene
is essentially conventional, with the effect of the pseudogap in the density
of states being adequately captured through the value of $g_{\ts}(\mu)$.

The remaining points in Fig.\ \ref{Fig7} represent $T_K^0$ for
$C_{3v}$-symmetric hollow-site adsorption, calculated assuming that
$g_{\hs}(E) = 6 [1 - J_0(2\sqrt{3}|E|/D)] \, g_{\ts}(E)$, a relation that holds
rigorously only for the region $|E|\ll D$ of linear dispersion. Although there
are doubtless some corrections to this relation within the energy range
$|E-\mu|\lesssim U$ that determines $T_K$ \cite{Haldane1978}, the approximation
is adequate to illustrate the qualitative differences between top- and
hollow-site adsorption. The circles in Fig.\ \ref{Fig7} correspond to the same
hybridization width $\Gamma=2$\ eV illustrated for top-site adsorption,
and show that $T_K^0$ is so small for $\mu \lesssim 0.26$\ eV  as to be
completely inaccessible to experiments, but then rises rapidly over a narrow
range of chemical potential so that for $\mu\gtrsim 0.28$\ eV, it exceeds the
Kondo scale for top site-adsorption. Over the entire range of $\mu$ covered in
the figure, the hollow-site $T_K^0$ exceeds by at least 50 orders of magnitude
the value predicted by Eq.\ \eqref{eq:T_K:Haldane}, pointing to strong
departures from conventional Kondo physics.

Based on the spatial geometries, we believe it likely that for a given adatom,
hollow-site adsorption will involve a smaller wave-function overlap between the
active impurity level and the $p_z$ orbital on any of the nearest carbon atoms
than would be the case for top-site adsorption. This suggests that the top-site
case $\Gamma=2$\ eV should more properly be compared with hollow-site adsorption
for some value $\Gamma<2$\ eV. The first-principles calculations required to
estimate the appropriate value of $\Gamma$ are beyond the scope of the present
work. However, the general idea can be seen by considering sample data for
$\Gamma=1.5$\ eV, plotted with squares in Fig.\ \ref{Fig7}. The $T_K^0$ vs $\mu$
curve for this case has the same shape as its hollow-site counterpart for
$\Gamma=2$\ eV, but it is shifted to higher $\mu$ values. Not surprisingly,
further reductions in $\Gamma$ lead to even larger shifts in the curve. 

If the hybridization matrix elements do not satisfy Eq.\ \eqref{eq:C_3v},
as will be the case for most $d$ and $f$ atomic orbitals, then following the
arguments in Ref.\ \onlinecite{Uchoa2011}, one should expect $\Gamma_{\hs}(E)$
for undeformed graphene instead to vanish linearly for $|E|\ll D$, with a
prefactor that depends on the degree of $C_{3v}$ symmetry breaking. In such
cases, the dependence of the Kondo scale on chemical potential should be very
similar to that shown in Fig.\ \ref{Fig7} for top-site adsorption, quite possibly
with a shift to the right arising from reduced hybridization matrix elements.

Results such as those shown in Fig.\ \ref{Fig7} suggest that experimental
observation of the Kondo effect for adatoms on undeformed graphene will depend
on the ability, via doping or application of back-gate voltages, to move the
chemical potential significantly (e.g., several hundred meV) away from the Dirac
point. Top-site and $C_{3v}$-symmetry-breaking hollow-site adsorption are
expected to display an exponential dependence of the Kondo temperature on the
value of the host LDOS at the chemical potential. Due to tunneling interference
effects, $C_{3v}$-symmetric hollow-site adsorption should exhibit an even
greater sensitivity to the location of the chemical potential.

\subsubsection{Hollow-site adsorption on deformed graphene}
\label{subsubsec:hollow:deformed}

Equation \eqref{eq:g_hs} remains valid in the presence of smooth
out-of-plane deformations of the graphene monolayer. As discussed in Secs.\
\ref{subsec:fields} and \ref{subsec:LDOS-changes}, such deformations modify
the continuum-limit electronic Green's function $G(\br, \br', E)$. Nearby
deformations are likely also to modify the hybridization matrix elements
$W_j$ between a hollow-site adatom and its surrounding carbon atoms. In
general, both the changes in the Green's function and those in the
hybridization matrix elements will break any $C_{3v}$ symmetry about the
impurity site that might have been present when the graphene was undistorted,
and can be expected to introduce into $g_{\hs}(E)$ terms proportional to
$\eta^2$ (the strain measure introduced in Sec.\ \ref{subsec:fields}) that
vanish at the Dirac points as $|E/D|$. Terms arising from changes in the
Green's function should reach their greatest magnitude at energies
$|E| = O(E_b)$, as is the case for top-site adsorption, while terms
originating in changes in hybridization matrix elements likely extend
throughout the energy range $|E|\lesssim D/6$ of Dirac dispersion. Since a
hollow-site adatom couples to both sublattices, deformations will have an
overall more muted impact on the Kondo temperatures than for top-site
adsorption and will not lead to distinctive alternating patterns analogous
to the ones described Sec.\ \ref{subsec:top}.

As shown in Sec.\ \ref{subsubsec:hollow:pristine}, the Kondo scale for
$C_{3v}$-symmetic hollow-site adsorption in pristine graphene varies almost
as a step function with respect to variation of the chemical potential,
rising over a very narrow window of $\mu$ from being undetectably small to
become larger than $T_K^0$ for top-site adsorption. Unless an experimental
system is fine-tuned into this window, the deformation-induced effects
discussed in the preceding paragraph will have negligible effect on the
Kondo temperature and on the prospects for experimental observation of Kondo
physics. For this reason, little purpose is served by performing detailed
numerical calculations for hollow-site adsorption in the presence of
deformation.

\section{Discussion}
\label{sec:discuss}

The theoretical and numerical work reported in this paper has investigated
factors that influence the characteristic temperature $T_K$ and energy scale
$k_B T_K$ of the Kondo effect for adatoms on graphene. The two-dimensional host
enters the Kondo physics through the hybridization function $g(E)$, which
provides a spectral description of adatom-host orbital overlaps. Depending on
the adsorption geometry, $g(E)$ for pristine graphene is expected to vanish with
an either linear or cubic dependence on $|E|$ on approach to the Dirac points at
$E=0$. As a result, $T_K$ shows strong sensitivity to the position of the
chemical potential $\mu$. For top-site adsorption of the magnetic atom directly
above a single host carbon, $T_K$ displays an exponential dependence on $\mu$
that is captured quite well by substituting the value $g(\mu)\propto|\mu|$ into
the standard expression for the Kondo scale in a conventional metal. Adsorption
of the magnetic atom in the hollow site in a high-symmetry configuration at the
center of a carbon hexagon, described by $g(E)\propto|E|^3$ for $|E|$ much
smaller than the half-bandwidth, yields a much sharper, almost step-like
variation of $T_K$ with increasing $\mu$. As a result, prospects of probing the
Kondo regime $T\lesssim T_K$ for hollow-site adsorption hinge on the ability to
dope or gate the chemical potential far from the Dirac points.

The main focus of the paper has been the exploration of strain as a tool
for enhancing the value of $T_K$ and revealing unique aspects of the Kondo
effect in graphene. We have shown that different placements of magnetic adatoms
relative to the peak of a slowly varying deformation yield wide variations in
the Kondo screening temperature with a spatial dependence that amplifies an
underlying pattern of strain-induced changes in the local density of states.
Fairly modest (smaller than 1\%) strains can locally increase the Kondo
temperature for a top-site adatom coupled to a single carbon atom from one
sublattice by at least an order of magnitude compared to the situation in
undeformed graphene, while simultaneously decreasing by a similar factor $T_K$
for nearby adsorption to the other sublattice. These effects can be observed
over a wide range of the model parameters $\veps_d$, $U$, and $\Gamma$
describing the adatom and its hybridization with the graphene host, and depend
crucially only on the chemical potential lying in the general energy range
where the LDOS on each sublattice is significantly affected by the deformation.
This unique pattern of spatial variation can be used as a fingerprint to
identify the Kondo regime for adatoms on graphene. Magnetic adatoms attached in
other geometries, such as the hollow-site configurations, are expected to
experience weaker strain-induced modulations in $T_K$.

In recent years, much progress has been achieved in the area of substrate
engineering for graphene \cite{YJiang2017,Zhang2018}. Setups like those reported
in Ref.\ \onlinecite{Zhang2018}, for example, create a periodic strain modulation
in graphene deposited on top of SiO$_2$ nanospheres. The weak graphene-substrate
hybridization in such experiments makes applicable the theoretical description
developed in this paper. Local probes, combined with atomic manipulation of
adatom placement, should allow observation of variations in $T_K$ that map
strain fields at a truly microscopic level.

\section{Acknowledgments}

We acknowledge support from NSF Grant Nos.\ DMR-1508325 (Ohio) and DMR-1508122
(Florida). D.Z. acknowledges support from the OU-CMSS Fellowship program. Portions of this work were completed at the Aspen Center for Physics
under support from NSF grant No.\ PHY-1607611.
---

\bibliography{strain}
\bibliographystyle{apsrev4-1}
\end{document}